\documentclass[10pt,superscriptaddress,twocolumn,amsmath,amssymb,aps,prl,showpacs]{revtex4-1}
\usepackage{mathrsfs}
\usepackage{graphicx}
\usepackage{dcolumn}
\usepackage{bm}
\usepackage{amssymb}
\usepackage{amsmath}
\usepackage{paralist}
\usepackage{float}
\usepackage{mdframed}
\usepackage{booktabs}
\usepackage{mathtools} 
\usepackage{graphicx}
\usepackage[colorlinks,linkcolor=blue,anchorcolor=blue,citecolor=green]{hyperref} 
\usepackage{cleveref}
\usepackage{url}
\usepackage{color}

\begin{document}

\title{Spin incoherent liquid  and interaction-driven criticality in 1D Hubbard model }

\author{Jia-Jia Luo}
\affiliation{State Key Laboratory of Magnetic Resonance and Atomic and Molecular Physics,
	Wuhan Institute of Physics and Mathematics,  APM, Chinese Academy of Sciences, Wuhan 430071, China}
\affiliation{University of Chinese Academy of Sciences, Beijing 100049, China.}

\author{Han Pu}
\email[]{hpu@rice.edu}
\affiliation{Department of Physics and Astronomy, and Rice Center for Quantum Materials,
Rice University, Houston, Texas 77251-1892, USA}

\author{Xi-Wen Guan}
\email[]{xiwen.guan@anu.edu.au}
\affiliation{State Key Laboratory of Magnetic Resonance and Atomic and Molecular Physics,
	Wuhan Institute of Physics and Mathematics, APM,  Chinese Academy of Sciences, Wuhan 430071, China}
\affiliation{NSFC-SPTP Peng Huanwu Center for Fundamental Theory, Xi'an 710127, China}
\affiliation{Department of Theoretical Physics, Research School of Physics and Engineering,
	Australian National University, Canberra ACT 0200, Australia}


\date{\today}

%

\begin{abstract}
Although the one dimensional (1D) repulsive Fermi-Hubbard model has been intensively studied over many decades, a rigorous understanding of many aspects of the model is still lacking. In this work, based on the solutions to the thermodynamic Bethe ansatz equations, we provide a rigorous study on the following: (1) We calculate the fractional excitations of the system in various phases, from which we identify the parameter regime featuring the spin incoherent Luttinger liquid (SILL). We investigate the universal properties and the asymprotic of correlation functions of the SILL. (2) We study the interaction-driven phase transition and the associated criticality, and build up an essential connection between the Contact susceptibilities and the variations of density, magnetization and entropy with respect to the interaction strength. As an application of these concepts, which hold true for higher dimensional systems, we propose a quantum cooling scheme based on the interaction-driven refrigeration cycle. 

 %
 %

\end{abstract}

\maketitle

One-dimensional (1D) Fermi-Hubbard model, describing strongly correlated electrons in a 1D lattice,  has become increasingly important in ultracold atoms, condensed matter and quantum metrology. 
Owing to the Bethe ansatz exact solution of the model \cite{Lieb:1968,Lieb:2003}, a variety of strongly correlated many-body phenomena have been extensively studied 
for over forty years, including Tomonaga-Luttinger liquid (TLL) \cite{Giamarchi:book,Imambekov:2012,Ess05},  spin-charge separation \cite{Boll:2016,Hilker:2017,Vijayan:2020,Spar:2022}, thermal and magnetic properties \cite{Krivnov:1975,Bogoliubov:1988,Lee:1988,Penc,Woynarovich:1983,Sacramento:1994,Essler:1994}, the Fulde-Ferrell-Larkin-Ovchinnikov (FFLO) \cite{Fulde:1964,Larkin:1965} pairing correlation \cite{Yang:2001,Tezuka:2008,Feiguin:2007,Kajala:2011,Cheng:2017A,Cheng:2017B}, etc. 
%
%
On the experimental side, 1D exactly solvable models have been successfully realized in the lab, allowing us to compare elegant and
sophisticated exact solutions directly with experimental measurements \cite{Cazalilla:2011,Guan:2013,Batchelor:2016,Mistakidis:2022}, 
Significant new experimental developments cover a broad range of physics such as 
the generalized hydrodynamics \cite{Kinoshita:2006,Langen:2015,Schemmer:2019}, dynamical fermionization \cite{Wilson:2020}, TLL \cite{Yang:2017,YangTL:2018,Meinert:2015}, fractional exclusion
 statistics \cite{Zhang:2022}, quantum holonomy \cite{Haller:2009,Kao:2021}, $p$-wave interacting fermions \cite{Chang:2020,Ahmed-Braun:2021,Jackson:2022}, high spin symmetry magnetism \cite{Pagano:2014,Song:2020}, etc.. For a recent review, see Ref. \cite{Guan:2022}.
%


Despite of these tremendous efforts, many aspects of the model still lack rigorous understanding. In particular, phase transitions have been extensively studied in the context of varying external potentials such as chemical potential and magnetic field. However, interaction-driven phase transitions have not received much attention even though interaction plays an essential role in many-body systems. This can be attributed to the fact that interaction strength is hardly tunable in traditional solid materials. The advent of cold atoms completely changed this situation as interaction strengths in atomic systems have been routinely controlled via Feshbach resonance. A notable recent example is the demonstration of the spin-charge separation \cite{Haldane:1981,Recati:PhysRevLett.90.020401,Guan:2012,Mestyan:2019spin,Patu:PhysRevB.101.035149} in a 1D continuum Fermi gas where the spin and the charge velocities are shown to exhibit distinct dependence on the interaction strength~\cite{Senaratne:2022}. 

Motivated by this, in this Letter, we show that the tunability of interaction strength allows further exploration of the spin incoherent Luttinger liquid (SILL) \cite{Fiete:2004,Fiete:2007,Cheianov:2004,Cavazos-Cavazos:2022} and interaction-driven quantum phase transitions in the Hubbard model. Specifically, we present rigorous results of fractional charge and spin excitations, analytical results on the asymptotic of single-particle Green's function and pair correlation functions of the SILL, and interaction-driven criticality. Furthermore, inspired by the notion of the partial wave Contact in ultracold Fermi gas \cite{Tan:2008,ZhangSZ:2009}, we build up general relations between Contact susceptibilities and the variation of density, magnetization and entropy with respect to the interaction strength, using which we propose a quantum cooling scheme based on the interaction-driven refrigeration cycle.

 \begin{figure}[th] 
\begin{center} 
	\includegraphics[width=0.99\linewidth]{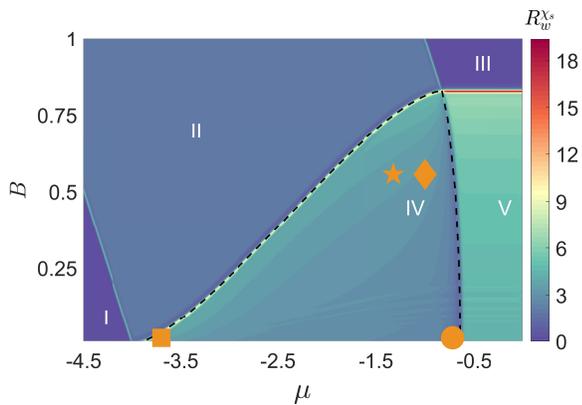} 
\end{center} 
\caption{Phase diagram represented by the contour plot of the Wilson ratio $R^{\chi_s}_w$ at temperature $T=0.005$ and $u=1$. The corresponding phases are: empty lattice I, partially filled and fully-polarized phase II, fully-filled and -polarized phase III, partially-filled and -polarized phase IV, and fully-filled and partially-polarized phase V (Mott insulator). The dotted lines represent analytic solution of BA equations obtained at zero temperature. The orange symbols indicates the locations of excitations plotted in Fig.~\ref{Fig2:excitation}.
}            
	\label{Fig1:phase}     
\end{figure}
 
 {\em 1D Hubbard model --} The 1D single-band Hubbard model is described by the Hamiltonian  \cite{Lieb:1968,Ess05} 
\begin{eqnarray}
H&=&- t \sum^L_{j=1, a=\uparrow,\downarrow} \left( c_{j,a}^\dagger c_{j+1,a} + {\rm h.c.}   \right) \nonumber\\
&& + u \sum_{j=1}^L \left(2n_{j,\uparrow}-1\right) \left(2 n_{j,\downarrow}-1\right)-\mu \hat{n} -2B\hat{S}^z,  \label{Ham}
\end{eqnarray}
where $c_{j,a}^\dagger$ ($c_{j,a}$) is the creation (annihilation) operator of an electron with
spin $a$ ($a=\uparrow$ or $\downarrow$) at site $j$ on a 1D lattice of length $L$, satisfying  the standard anticommutation relations.
$t$, $\mu$ and $B$ are hopping amplitude, chemical potential and magnetic field, respectively. In this work, we will only consider repulsive interaction with $u>0$ and take $t=1$ as the unit of the system. $\mu$ and $B$ are renormalized accordingly and become dimensionless. 
Meanwhile $n_{j,a}=c_{j,a}^\dagger c_{j,a}$  and $\hat{n}=\frac{\hat{N}}{L}=\frac{1}{L}\sum_{j,a} n_{j,a}$ are the density operator and the average fermion number per lattice site, respectively. We denote the magnetization $\hat{S}^z=\sum_{j,a}\left( n_{j,\uparrow }- n_{j,\downarrow }\right)$.
For vanishing external potentials ($\mu=0,\, B=0$) and even $L$, Hamiltonian (\ref{Ham}) possesses $SO(4)\cong SU(2) \times SU(2)/\mathbb{Z}_2$ symmetry,  preserving spin rotational and $\eta$-pairing symmetries \cite{Ess05,YangCN:1989,YangCN:1990}, i.e. 
$ {[H,S^{\alpha}]}=0= 
[H,\eta^{\alpha}]$ with $\alpha=x,y,z $. The spin and $\eta$-pair operators can be transformed to each other via Shiba transformation, showing the connection between spin and charge \cite{Ess05}, also see Supplementary Materials \cite{SM}.
We will use spin and $\eta$-pair  magnetizations $S^z$, $\eta^z=\frac{1}{2}\left(N-L\right)$ to characterize the fractional spin and charge excitations. 

In 1968 Lieb and Wu \cite{Lieb:1968} derived the BA equations for the 1D Hubbard  model by means  of Bethe's hypothesis \cite{Bethe:1931}. 
Takahashi \cite{Takahashi:1972} later found the root patterns of the BA equations, i.e. real $k$, length-$n$ $\Lambda$ string (known as spinon bound state) composed of $n$ spin-down electrons, length-$m$ $k$-$\Lambda$ string containing $m$ down-spin and $m$ up-spin particles, which determine both the ground and the excited states of the model. 
%
Building on Takahashi's string hypothesis, 
and using the Yang-Yang method \cite{YangCN:1969}, one can obtain the thermodynamic Bethe ansatz (TBA) equations of the model \cite{Takahashi:1972} (for convenience, see \cite{SM}). 
%
In principle, all thermodynamic properties of the model can be obtained from the TBA. 
However, solving the infinite number of nonlinear integral TBA equations poses a tremendous theoretical challenge. 
Therefore many important questions remain to be answered. 

\begin{figure}[h] 
	\begin{center} 
	\includegraphics[width=1.0\linewidth]{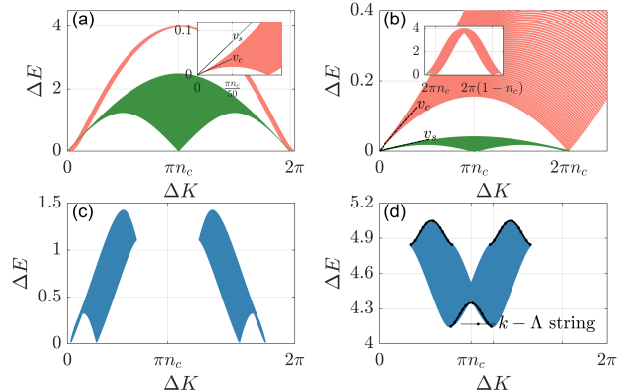} 
	\end{center} 
	\caption{Elementary fractional spin and charge excitations. The orange areas in (a) and (b) represent particle-hole excitations of charge, whereas the green parts show the two-spinon  excitations with quantum numbers $\left(\Delta\eta^z,\Delta S^z\right)=(0,1)$ induced from spin flipping. (c) Fractional antiholon-spinon excitations $(\frac{1}{2},-\frac{1}{2})$ , i.e. adding an extra spin-down electron  to create an antiholon and a spinon.  (d) Gapped excitation spectra for length-1 $k$-$\Lambda$ string and a length-2 $\Lambda$ string. All graphs are drawn in the first Brillouin zone with interaction $u=1$ and the parameters (a): $B=0,\mu=-0.6619$, density $n=0.9801$ (near half-filled band); (b): $B=0,\mu=-3.8508$, density $n=0.1389$ (dilute limit); (c): $B=0.555,\mu=-1$; (d): $B=0.555,\mu=-1.32$.}
\label{Fig2:excitation}            
\end{figure}

{\em Low-temperature phase diagram and fractional excitations --}  
A rich phase diagram of the 1D Hubbard model (\ref{Ham}) in magnetic field-chemical potential plane can be obtained from either the BA or the TBA equations at zero temperature. We find that the dimensionless Wilson ratio (WR) $R^{\chi_s}_w=\frac{4}{3}\left(\frac{\pi k_B}{\mu_Bg}\right)^2\frac{\chi_s}{C_v/T}$, where $\chi_s$ is the spin susceptibility and $C_v$ the specific heat, conveniently characterizes the TLLs.
Here $k_B$, $\mu_B$ and $g$ are the Boltzmann constant, the Bohr magneton and the Land\'{e} factor, respectively, which we set to be unity in our calculation. 
The value of the WR is temperature independent at low energy and exhibits a sudden change in the vicinities of the phase boundaries. 
This feature naturally maps out the full phase diagram of the 1D repulsive Hubbard model, as we show in Fig.~\ref{Fig1:phase}. 
Specifically, we observe that the values of the WR in Fig.~\ref{Fig1:phase} confirm the bosonization result \cite{Fiete:2007}, i.e. $R^{\chi_s}_w=2v_cK_s/(v_s+v_c), \, 2$ and  $4K_s$  for the TLLs in Phase IV, II and V, respectively, where $K_s$ is the Luttinger parameter for spin, $v_{c,s}$ are sound velocities for charge and spin, respectively. 
The WR is zero for Phases I and III, see SM \cite{SM} for more details.

Figure~\ref{Fig2:excitation} (a) and (b) demonstrate excitations at zero magnetic field in charge and spin degrees of freedom near the half-filled lattice and the dilute limit, respectively. 
The particle-hole excitation of charge (orange) forms continuum spectra within the first Brillouin zone.
Flipping one spin leads to excitation spectrum (green) of two deconfined spinons  with a fractional spin-$\frac{1}{2}$.
 In the long wavelength limit, i.e. $\Delta K\to 0$, both charge and spin excitations exhibit linear dispersion: $ \Delta E_{c,s} = v_{c,s} \hbar |\Delta K|$, where $v_c=0.5995,v_s=1.2403$ in  (a) and  $v_c=0.7206,v_s=0.1454$ in (b), showing spin and charge separated excitations. 
 However, subtle differences between these two limits are observed, i.e. for (a), the charge excitation displays a single-particle nature due to the vanishing of the charge Fermi sea; 
 for (b), the spin and charge excitations are significantly separated, making this preferred region to observe
spin-charge separation \cite{SM}.
Later, we will further demonstrate the existence of the SILL in region (b) for temperature $E_s\ll k_BT\ll E_F$, where $E_{s,c}\sim k_Fv_{s,c} $ with $k_F=\pi n_c$ are the spin and charge energies, respectively.  
Fig.~\ref{Fig2:excitation}(c) shows the fractional antiholon-spinon  excitation spectra with the $\eta$-pair and spin magnization $(\Delta \eta^z, \Delta S^z)=(\frac{1}{2},-\frac{1}{2})$ by adding an antiholon particle $N_e=N+1$ superposed with one spinon particle in $M_1$ sector, which are outside of the spin-charge separated TLL regime \cite{SM}.
Fig.~\ref{Fig2:excitation}(d) shows the two gapped excitations, i.e., length-$1$ $k$-$\Lambda$ string and length-$2$ $\Lambda $ spinon bound states, forming a gapped continuum band.

{\em Universal scaling laws, SILL, correlation functions --} 
Rigorous results on quantum criticality of the repulsive Hubbard model remain largely unknown. At zero temperature, the phase transition occurs at a quantum critical point (QCP) where a degree of freedom appears, disappears or reaches saturation.
At finite temperature, the QCP fans out into the V-shaped quantum critical regime, in which the free energy takes universal form. By considering the relevant degrees of freedom, we can simplify the TBA equations and find such universal forms. In the SM~\cite{SM}, we have derived analytically the free energies of all quantum critical regions associated with various phase transitions of the 1D Hubbard model. Here, we only write down the free energy 
at quantum criticality for the II-IV transition:
{\small\begin{eqnarray}
f=f_0-\frac{\pi T^2}{6v_c}+T^{\frac{3}{2}}\pi^{\frac{1}{2}}\sigma_1(0)\left(\frac{\varepsilon^{''}_1(0)}{2}\right)^{-\frac{1}{2}}\mathrm{Li}_{\frac{3}{2}}\left(-\mathrm{e}^{\frac{-\varepsilon_1(0)}{T}}\right),
\label{24-45}
\end{eqnarray}}
where $\mathrm{Li}_n$ denotes the polylog functions, $f_0$ is the ground-state energy, the $T^2$ terms represents the contributions from the collective excitations of the background degree of freedom near the QCP,
$\sigma_1(0)$ denotes the spin density at $\Lambda=0$,  $\varepsilon_1(\Lambda)$ is the dressed energy of length-$1$ string,  and $\varepsilon^{''}_1(0) \equiv \left. \frac{d^2\varepsilon_1}{d\Lambda^2} \right|_{\Lambda=0}$, 
with 
\begin{eqnarray}
\varepsilon_1(0)&=-&\alpha_B\Delta B-\alpha_\mu\Delta \mu-\alpha_u\Delta u,
\label{alpha-beta}
\end{eqnarray}
denoting the spin dressed energy gap away from the QCP,
where $\Delta B=B-B_c,\,\Delta \mu=\mu-\mu_c,\,\Delta u=u-u_c$ are distances away from the QCP ($B_c,\, \mu_c,\, u_c$).
The analytic expressions of the factors $\alpha_{B,\mu,u}$ are rather cumbersome and can be found in the SM \cite{SM}. As we will show below,
the free energy in Eq.~(\ref{24-45}) elegantly leads to and provides a rigorous understanding of the universal thermodynamic properties of the TLL, the SILL and the quantum scaling laws at criticality. 

\begin{figure}[h] 
	\begin{center} 
	\includegraphics[width=0.99\linewidth]{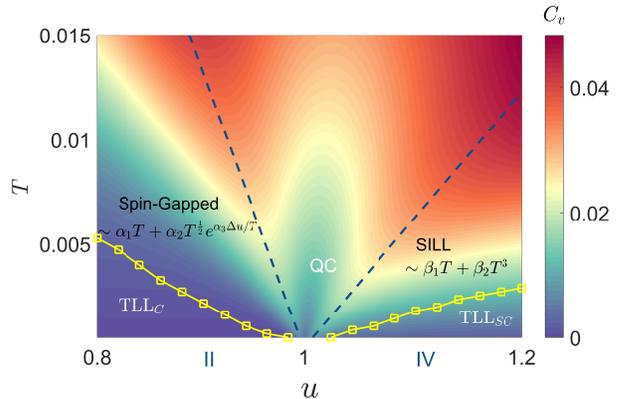} 
	\end{center}   
	\caption{Contour plot of specific heat in the $T$-$u$ plane at $\mu=-2,B=0.55$ for II-IV transition. The blue dashed lines present the critical temperatures determined by the maximum values of the specific heat (Eq.(\ref{u-cv})). The yellow lines with square symbols mark the TLL phase boundary, below which the specific heat shows a linear temperature dependence. Crossover regimes between the blue dashed and the yellow lines denote the spin-gapped phase (on the left) and the SILL phase (on the right).}
\label{Fig3:Cv-u}
\end{figure} 

 In the quantum critical regime, $T\gg \Delta u$, and from Eq.~(\ref{24-45}), the specific heat can be readily derived as
\begin{eqnarray}
C_v/T &=&  c_0+c_1T^{-1/2} \left[ \frac{3}{4} \mathrm{Li}_{\frac{3}{2}}\left(-\mathrm{e}^x\right)- x\mathrm{Li}_{\frac{1}{2}}\left(-\mathrm{e}^x\right)\right.\nonumber\\
&&\left. + x^2\mathrm{Li}_{-\frac{1}{2}}\left(-\mathrm{e}^x\right)\right]+O((\Delta u/T)^{5/2}),\label{u-cv}
\end{eqnarray} 
where $x\equiv \alpha_u \Delta u/T$, and $c_{0,1}$ denote the regular part and a constant depending on the critical point $u_c$, respectively.   
In Fig.~\ref{Fig3:Cv-u}, we display the contour plot of the specific heat in the plane spanned by $T$ and $u$ at $\mu=-2,\,B=0.55$. The specific heat shows a bimodal structure, whose local maxima mark the critical temperatures. The local maxima can be determined by $\partial C_v/\partial u=0$, leading to $x_1=-1.5629$ and $x_2=3.6205$, corresponding to the two blue dashed lines in Fig.~\ref{Fig3:Cv-u}. These two lines join at $u_c=1$ at $T=0$ and the quantum critical regime resides between them, displaying a universal free fermion criticality,  i.e. dynamical and correlation critical exponents $Z=2,\,\nu=1/2$, respectively.

The yellow line with square symbols in Fig.~\ref{Fig3:Cv-u} represents the boundary of the TLL region which lies below the line. In the TLL, $C_v$ is linear in $T$. To the left of $u_c$, the TLL (labelled as TLL$_C$) contains only the charge component and $C_v=\frac{\pi}{3}\frac{T}{v_c}$. To the right of $u_c$, the TLL (label as TLL$_{SC}$) contains both the spin and the charge component with $C_v=\frac{\pi}{3}\left( \frac{1}{v_c}+\frac{1}{v_s} \right)T $. This is the regime where spin-charge separation \cite{Haldane:1981,Recati:PhysRevLett.90.020401,Guan:2012,Mestyan:2019spin,Patu:PhysRevB.101.035149} can be observed. 
 Above TLL$_C$, the spin sector is gapped and the specific heat is given by $C_v\approx \alpha_1 T+\alpha_2 T^{1/2}\mathrm{e}^{\alpha_3\Delta u/T}$, where $\alpha_{1,\,2,\,3}$ are constants. By contrast, above TLL$_{SC}$, we find a region of SILL in the temperature range given by $k_Fv_s\ll  k_BT\ll k_Fv_c$, exhibiting a propagating charge mode but not a spin mode with the corresponding specific heat $C_v\approx \beta_1 T+\beta_2 T^3$ with $\beta_{1,2}$ being constants \cite{SM}, showing a gas-liquid co-existence in the SILL. 

In the SILL regime, the spin excitation is suppressed and hence the spin sector is non-dynamic, while charge maintains relevant at low energy. Taking the reasonable limits $|x\pm iv_ct|\ll v_c/T$ and $|x\pm iv_st|\gg v_s/T$, we can calculate the finite-temperature single-particle Green's function and pair correlation function
\begin{eqnarray}
G^{\uparrow }&\! \approx \! & \mathrm{e}^{-\mathrm{i} k_{F,\uparrow} x}{\cal C}_{\uparrow}^-\left( x-\mathrm{i} v_c t\right) \langle S_R^+\left(x,t\right)S_R(0,0)\rangle
+h.c.,\nonumber \label{Correlation-Green}\\
G^{p}&\! \approx \! & \mathrm{e}^{-\mathrm{i} \left( k_{F,\uparrow} + k_{F,\downarrow} \right) x}{\cal C}_{p}^-\left( x-\mathrm{i} v_c t\right) 
{\cal C}_{p}^+\left( x+\mathrm{i} v_c t\right)\nonumber \\
&& \times \langle S_R^+\left(x,t\right)S_R(0,0)\rangle +h.c.,  \label{Correlation-pair}
\end{eqnarray}
where the charge correlations ${\cal C}_{\uparrow}^-\left(Z\right)\sim  1/Z^{2\Delta_c^+}$, ${\cal C}_{p}^{\pm} \left(Z\right)\sim  1/Z^{2\Delta_c^{\mp}}$ decay as a power-law of distance, whereas the spin mode correlation $\langle S_R^+\left(x,t\right)S_R(0,0)\rangle\sim \left(2\pi \alpha k_F\right)^{2\Delta^{+}_s+2\Delta^{-}_s}\mathrm{e}^{-\pi\alpha\left(2\Delta^{+}_s+2\Delta^{-}_s\right) k_Fx}$ decays exponentially. Here $\Delta_c^{\pm}$ are the conformal dimensions which can be calculated analytically and numerically \cite{SM}, $\alpha$ is a constant. For the particular case $B=0$, our results agree with those given in Ref.~\cite{Cheianov:2004}. We comment that the SILL has been theoretically studied under the framework of bosonization \cite{Fiete:2004,Fiete:2007,Cheianov:2004}. Our work here provides a rigorous underpinning of the SILL based on the TBA.

\begin{figure}[h] 
	\begin{center} 
	\includegraphics[width=0.99\linewidth]{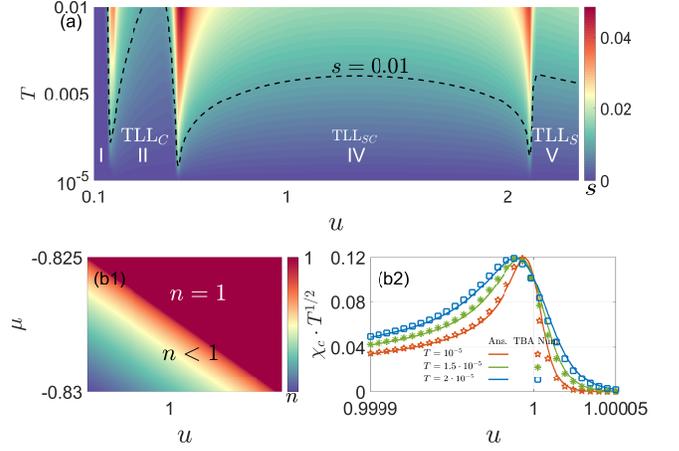} 
	\end{center}       
	\caption{(a) Contour plot of the entropy in $T-u$ plane for  $B=0.15,\,\mu=-2.5$. Black dotted curve is the isentropic line for $s=0.01$. When the interaction increases, the system enters sequentially into Phases II, IV and V. (b1) Contour plot of density $n$ near the IV-V phase transition with  $u=1,\,\mu=-0.82724$ and $B_c=0.82714$. (b2) Scaling behaviour of charge susceptibilities near phase transition from IV to V driven by interaction.} \label{Fig4:sn}
\end{figure}

{\em Contact susceptibilities and quantum cooling --}  
In analogy to the Contact for quantum gases \cite{Tan:2008,ZhangSZ:2009}, here we define the lattice version of the Contact $C=\partial f/ \partial u=4 d -2n+1$, where $n=\langle \hat{n} \rangle$, and $d=\frac{1}{N}\sum_i\left\langle n_{i,\uparrow}n_{i,\downarrow} \right\rangle$ is the average double occupancy, a quantity which can also depict the phase diagram \cite{SM}. It is, however, more essential to define Contact susceptibilities with respect to the external potentials. Using the Maxwell relations, we may build up general relations between Contact susceptibilities and interaction-driven variations of entropy, density and magnetization \cite{SM}:
\begin{equation}
    \frac{\partial s}{\partial u}=-\frac{\partial C}{\partial T},\;\;\;
\frac{\partial n}{\partial u}=-\frac{\partial C}{\partial \mu},\;\;\;
\frac{\partial m}{\partial u}=-\frac{\partial C}{\partial (2B)}. \label{uT}
\end{equation}
These relations provide deep insights into the  interaction effects and universal behaviour of phase transitions. 

As a specific example, we now use the first relation in Eqs.~(\ref{uT}) to investigate interaction-driven quantum cooling. 
 Fig.~\ref{Fig4:sn} (a) shows a contour plot of entropy in the $T$-$u$ plane for fixed $B$ and $\mu$. The interaction-driven phase transitions from I to II, II to IV, and IV to V occur sequentially with increasing interaction strength. We observe a single-component charge TLL$_C$ in II, a spin and charge separated TLL$_{SC}$ in IV, and a spin TLL$_S$ in the Mott phase V. Conducting the total derivative of entropy with respect to the interaction $u$, the phase points on the isentropic line in the $T$-$u$ plane admit the relation
$
\frac{C_v}{T} \frac{\partial T}{\partial u}=\frac{\partial C}{\partial T}.
$
Thus the interaction-driven Gr{\"u}neisen parameter \cite{Yu:2020} defined by  $\Gamma_{int}=\frac{u}{C_v}\frac{\partial C}{\partial T}$ quantifies the efficiency of interaction-driven refrigeration. Near a critical point, local maximum of the entropy leads to a local temperature minimum in an isentropic process, and using the condition $\frac{\partial C}{\partial T}=0$, we have 
 $ \frac{1}{2}\mathrm{Li}_{\frac{1}{2}}\left(-\mathrm{e}^{x}\right)-x\mathrm{Li}_{-\frac{1}{2}}\left(-\mathrm{e}^{x}\right)=0$
that gives  a general solution $x\equiv: \alpha_u \Delta u/T\approx1.3117 $.
 Using the free energy Eq.~(\ref{24-45}), we can obtain the explicit expression of the maximum entropy near the transition point from II to IV $s_{c1}\approx \lambda_1 \pi^{1/2}\sigma_1(0)(\varepsilon^{''}_1(0)/2)^{-1/2}T_{c1}^{1/2}$, where  $\lambda_1=x\mathrm{Li}_{1/2}\left(-\mathrm{e}^{x}\right)-3/2\mathrm{Li}_{3/2}\left(-\mathrm{e}^{x}\right)\approx 1.3467$. 
 Similarly, for phase transition from  the Mott phase V to the phase IV, the maximum entropy is given by $s_{c2}\approx \lambda_1 \pi^{1/2}\rho(\pi)(-\kappa^{''}(\pi)/2)^{-1/2}T_{c2}^{1/2}$, where $\rho(\pi)$ denotes the charge density at $k=\pi$ and $\kappa^{''}(\pi) \equiv \left. \frac{d^2\kappa}{dk^2} \right|_{k=\pi}$, with $\kappa(k)$ the charge dressed energy \cite{SM}.
 On the other hand, the entropy in the Luttinger liquid phases TTL$_C$ and TTL$_S$ are given by $s_{L1}=\pi T_{L1}/{3v_c}$ and $s_{L2}=\pi T_{L2}/{3v_s}$, respectively. 
Therefore through an interaction-driven refrigeration cycle near phase transitions from II to IV and from V to IV in the $T$-$u$ plane, we can show that the reachable minimum temperatures are given by 
\begin{eqnarray}
	\frac{T_{c1}^{1/2}}{ T_{L1}}&=&\frac{\pi^{1/2}(\varepsilon^{''}_1(0)/2)^{1/2}}{3\lambda_1v_c\sigma_1(0)}, \\
	\frac{T_{c2}^{1/2}}{ T_{L2}}&=&\frac{\pi^{1/2}(-\kappa^{''}(\pi)/2)^{1/2}}{3\lambda_1v_s\rho(\pi)},
\end{eqnarray}
respectively. The minimum temperature in the $T$-$u$ plane is governed by  the  relation  $\alpha_u \Delta u/T\approx1.3117$ \cite{SM}. We remark that efficient cooling in lattice is a significant experimental challenge in ultracold atomic gases, the lack of which poses as a roadblock for realizing some exotic quantum phases. 

On the other hand, the other two relations in Eqs.~(\ref{uT}) provide essential insights for charge (IV-V) and spin (II-IV) phase transitions, respectively. 
Using these, we find two useful relations among the parameters $\alpha_{u,\mu,B} $  in Eq. (\ref{alpha-beta})
\begin{equation}
\frac{\alpha_u}{\alpha_{\mu}}=-\frac{\partial \mu}{\partial u},\quad \frac{\alpha_u}{\alpha_B}=-\frac{\partial B}{\partial u}, \label{mu}
\end{equation}
that provide us deep insights into the quantum criticality driven by dynamical interaction and external poentials. 
For example, Fig.~\ref{Fig4:sn}(b1) shows the phase transition from Phase IV to the Mott phase V in the $\mu-u$ plane,  
where $\frac{\partial \mu}{\partial u}$ is the slope along the transition line $n=1$. 
From Eq. (\ref{mu}) with fixed $B_c=0.82714$ around $u_c=1,\,\mu_c=-0.8272$ in Fig.~\ref{Fig4:sn}(b1), we may numerically get $\alpha_u \approx-1.9627$. With this and using the scaling form of  free energy given in  \cite{SM}, we can obtain the scaling behaviour of the compressibility in terms of $u$, which is in excellent agreement with numerical calculation from TBA equations, see  Fig.~\ref{Fig4:sn}(b2).

{\em Summary --}  We have presented new rigorous results of the 1D repulsive Hubbard model. We focus on the interaction-driven quantum criticality which has been largely ignored in previous studies. We studied the fractional excitations from which the SILL regime is identified and carefully investigated. We introduced several Contact susceptibilities and show how they provide crucial new insights into the system. Finally, we proposed a quantum cooling scheme based on the interaction-driven refrigeration cycle, which can potentially open up new avenues of research in reaching unprecedented low temperatures in lattice quantum gases. We note that some of the key concepts developed here will hold true in higher dimensions.   

\section*{Acknowledgement}
J.J.L and X.W.G. is supported by the NSFC key grant No. 12134015, the NSFC grant  No. 11874393 and No. 12121004. H.P. acknowledges support from the US NSF (PHY-2207283) and the Welch Foundation (Grant No. C-1669).



\clearpage\newpage
\setcounter{figure}{0}
\setcounter{table}{0}
\setcounter{equation}{0}
\def\thefigure{S\arabic{figure}}
\def\thetable{S\arabic{table}}
\def\theequation{S\arabic{equation}}
\setcounter{page}{1}
\pagestyle{plain}

\begin{widetext}

\section*{Supplementary material: Spin incoherent liquid  and interaction-driven criticality in 1D Hubbard model}
\begin{center}
{Jia-Jia Luo, Han Pu, Xi-Wen Guan}
\end{center}

\section{The Hubbard model}

The Hamiltonian of 1D Hubbard model is
\begin{equation}
H=-t\sum_{j=1}^{L} \sum_{a=\uparrow \downarrow}\left(c_{j, a}^{\dagger} c_{j+1, a}+c_{j+1, a}^{\dagger} c_{j, a}\right)+u \sum_{j=1}^{L} (1-2n_{j \uparrow}) (1-2n_{j \downarrow})-\mu\hat{N}-2B\hat{S}_z,\label{Ham}
\end{equation}
where $c_{j, a}^{\dagger}$ and $c_{j, a}$ are creation and annihilation operators of electrons with spin $a(\uparrow,\downarrow)$ situated in site $j$ on a 1D lattice of length $L$. $t$, $\mu$ and $B$ are hopping amplitude, chemical potential and magnetic field, respectively. 
$n_{j a}=c_{j,a}^\dagger c_{j,a}$ is particle density operator, $\hat{N}=\sum_{j=1}^{L}(n_{j \uparrow}+n_{j \downarrow})$ is the total number particle operator, and $\hat{S}_z=\frac{1}{2}\sum_{j=1}^{L}(n_{j \uparrow}-n_{j \downarrow})$ is the magnetization operator. 
$u$ denotes interaction strength and here, we focus on repulsive case $u>0$. 
In later analysis  the hopping strength $t$ and $k_B,\, h$ are  set to unity.
Thus, all thermodynamic quantities are dimensionless in all calculations.

For vanishing external potentials ($\mu=0,\, B=0$) and even lattice sites, Hamiltonian (\ref{Ham}) possesses $SO(4)\cong SU(2) \times SU(2)/\mathbb{Z}_2$ symmetry,  preserving spin rotational and $\eta$-pairing symmetries, i.e. 
$ {[H,S^{\alpha}]}=0$, 
$[H,\eta^{\alpha}]=0$ with $\alpha=x,y,z $, where the spin and $\eta$-pair operators are given in Ref.~\cite{Ess05},
\begin{eqnarray}
S^{\alpha}&=&\frac{1}{2}\sum_{j=1}^L\sum_{a,b=\uparrow,\downarrow}c_{j, a}^{\dagger}(\sigma^{\alpha})|_b^ac_{j,b},\\
\eta^x&=&-\frac{1}{2}\sum_{j=1}^L(-1)^j\left(c_{j, \uparrow}^{\dagger}c_{j, \downarrow}^{\dagger}+c_{j, \uparrow}c_{j, \downarrow}\right),\\
\eta^y&=&\frac{i}{2}\sum_{j=1}^L(-1)^j\left(c_{j, \uparrow}^{\dagger}c_{j, \downarrow}^{\dagger}-c_{j, \uparrow}c_{j, \downarrow}\right),\\
\eta^z&=&\frac{1}{2}\left(N-L\right).
\end{eqnarray}
In terms of these two discrete symmetries, the spin and the charge degrees of freedom can be transformed into each other in some ways, rendering fractional spin and charge excitations characterized by $S^z$ and $\eta^z$.

The Bethe ansatz equations for the 1D Hubbard  model was derived by Lieb and Wu \cite{Lieb:1968}. 
Here we write down explicitly the Bethe ansastz equations in terms of Takahashi \cite{Takahashi:1972} 
the root patterns, i.e. real $k$, length-$n$ $\Lambda$ string composed of $n$ spin-down electrons, length-$m$ $k$-$\Lambda$ string containing $m$ down-spin and $m$ up-spin particles.
As being introduced in the main text,  let us denote $M_e,\, M_n,\, M^{\prime}_n$  respectively the numbers of real $k$, length-$n$ $\Lambda$ and length-$n$ $k-\Lambda$ strings, thus the total particle number $N$ and spin down electron number $M$ are given by \cite{Ess05} $N=\mathcal{M}_{e}+\sum_{n=1}^{\infty} 2 n M_{n}^{\prime}$ and 
$M=\sum_{n=1}^{\infty} n\left(M_{n}+M_{n}^{\prime}\right)$, respectively. 
We further denote $\rho^p,\sigma_{n}^p,\sigma_{n}^{\prime p}$ ($\rho^h,\sigma_{n}^h,\sigma_{n}^{\prime h}$) as the densities for particles (holes) of $k,\, \Lambda,k-\Lambda$ sector, the root distributions of these types of strings are given by \cite{Takahashi:1972}
\begin{eqnarray}
\rho^p(k)+\rho^h(k)&=&\frac{1}{2\pi}+\cos k \sum_{n=1}^{\infty} \int_{-\infty}^{\infty} \mathrm{d} \Lambda a_{n}(\Lambda-\sin k)\left(\sigma_{n}^{\prime p}(\Lambda)+\sigma_{n}^p(\Lambda)\right), \\
\sigma_{n}^h(\Lambda) &=& -\sum_{m=1}^{\infty} A_{n m} * \sigma_{m}^p(\Lambda) +\int_{-\pi}^{\pi} \mathrm{d} k a_{n}(\sin k-\Lambda) \rho^p(k),
\\
\sigma_{n}^{\prime h}(\Lambda) &=&\frac{1}{\pi} \operatorname{Re} \frac{1}{\sqrt{1-(\Lambda-i n u)^{2}}} -\sum_{m=1}^{\infty} A_{n m} * \sigma_{m}^{\prime p}(\Lambda) -\int_{-\pi}^{\pi} \mathrm{d} k a_{n}(\sin k-\Lambda) \rho^p(k), \label{eq:density} 
\end{eqnarray}
where $a_n(x)=\frac{1}{2\pi}\frac{2nu}{(nu)^2+x^2}$, and  *  stands for the convolution
\begin{equation}
A_{nm}*f|_x=\int_{-\infty}^{\infty} \frac{\mathrm{d} y}{2 \pi}\frac{\mathrm{d} }{\mathrm{d}x}\Theta_{nm}\left(\frac{x-y}{u}\right)f(y).
\end{equation}
The derivation of the function $\Theta_{nm}$ reads 
{\small\begin{equation}
\frac{1}{2 \pi}\frac{\mathrm{d} }{\mathrm{d}x}\Theta_{nm}\left(\frac{x-y}{u}\right)=\left\{\begin{array}{l}
a_{|n-m|}\left(x-y\right)+2 a_{|n-m|+2}\left(x-y\right)+\cdots+2 a_{n+m-2}\left(x-y\right)+a_{n+m}\left(x-y\right), \text { if } n \neq m \\
2 a_{2}\left(x-y\right)+2 a_{4}\left(x-y\right)+\cdots+2 a_{2n-2}\left(x-y\right)+a_{2n}\left(x-y\right), \text { if } n=m
\end{array}\right..
\end{equation}}
The above equations can be used to study ground state properties and excitations in spin and charge. 

Building on the above Takahashi string solutions \cite{Takahashi:1972}, finite temperature thermodynamics of the model can be obtained from  the TBA equations in terms of the  dressed energies $\kappa(k),\varepsilon_n(\Lambda),\varepsilon_n^{\prime}(\Lambda)$, associated with the product of the logarithm of the ratio of hole density to particle density and temperature $\kappa=T\ln\left(\frac{\rho^h}{\rho^p}\right),\varepsilon_n=T\ln\left(\frac{\sigma_n^h}{\sigma_n^p}\right),\varepsilon_n^{\prime}=T\ln\left(\frac{\sigma_n^{\prime h}}{\sigma_n^{\prime p}}\right)$, namely, explicit TBA equations are given by
\begin{eqnarray}
\kappa(k)&=&-2 \cos k-\mu-2 u-B+\sum_{n=1}^{\infty} \int_{-\infty}^{\infty} \mathrm{d} \Lambda a_{n}(\sin k-\Lambda) T\ln \left(1+\mathrm{e}^{-\frac{\varepsilon_n^{\prime}(\Lambda)}{T}}\right)\\
&&-\sum_{n=1}^{\infty} \int_{-\infty}^{\infty} \mathrm{d} \Lambda a_{n}(\sin k-\Lambda) \ln \left(1+\mathrm{e}^{-\frac{\varepsilon_n(\Lambda)}{T}}\right),\nonumber \\
\varepsilon_n(\Lambda)&=&2nB-\int_{-\pi}^{\pi} \mathrm{d} k \cos k a_{n}(\sin k-\Lambda) T\ln \left(1+\mathrm{e}^{-\frac{\kappa(k)}{T}}\right)+\sum_{m=1}^{\infty} A_{n m} * T\ln \left(1+\mathrm{e}^{-\frac{\varepsilon_m(\Lambda)}{T}}\right), \\
\varepsilon_n^{\prime}(\Lambda)&=&4 \textmd{Re} \sqrt{1-(\Lambda-i n u)^{2}}-2 n \mu-4 n u-\int_{-\pi}^{\pi} \mathrm{d} k \cos k a_{n}(\sin k-\Lambda) T\ln \left(1+\mathrm{e}^{-\frac{\kappa(k)}{T}}\right)\\&&+\sum_{m=1}^{\infty} A_{n m} * T\ln \left(1+\mathrm{e}^{-\frac{\varepsilon_m^{\prime}(\Lambda)}{T}}\right).\nonumber\label{eq:TBA}
\end{eqnarray}
The Gibbs free energy per site is given in terms of the dressed energies
\begin{eqnarray}
f&=&-T \int_{-\pi}^{\pi} \frac{\mathrm{d} k}{2 \pi} \ln \left(1+\mathrm{e}^{-\frac{\kappa(k)}{T}}\right)+u\nonumber \\
&&-T \sum_{n=1}^{\infty} \int_{-\infty}^{\infty} \frac{\mathrm{d} \Lambda}{\pi} \operatorname{Re} \frac{1}{\sqrt{1-(\Lambda-i n u)^{2}}}\ln \left(1+\mathrm{e}^{-\frac{\varepsilon_n^{\prime}(\Lambda)}{T}}\right). \label{eq:f_t} 
\end{eqnarray}
From Eq.~(\ref{eq:f_t}), thermal and magnetic properties of the model can be obtained by using the standard statistical relations.

In the limit of zero temperature, the dressed energy of bulk kinds and lengths of strings is always positive in the parameter space. 
Consequently, they offer  no contributions to the integrals due to the logarithm function in the integral kernel. 
The only survivals are $k$ and length-1 $\Lambda$ string, and the values of  arguments $k,\Lambda$ should be cut off at points where dressed energies switch the sign. Thereby, coupled equations for densities and dressed energies\cite{Lieb:1968} are greatly simplified
\begin{eqnarray}
\kappa(k)&=&-2 \cos{k}-\mu-2 u-B+\int_{-A}^{A}\mathrm{d} \Lambda  a_1(\sin{k}-\Lambda) \varepsilon_1(\Lambda), \label{kappa}\\
\varepsilon_1(\Lambda)&=&2 B +\int_{-Q}^{Q} \mathrm{d} k \cos k a_1(\sin{k}-\Lambda) \kappa(k)
-\int_{-A}^{A} \mathrm{d} \Lambda^{\prime} a_{2}\left(\Lambda-\Lambda^{\prime}\right)\varepsilon_1(\Lambda^{\prime}). \label{epsilon}
\end{eqnarray}
And total quantities $\rho=\rho^p+\rho^k,\sigma_{1}=\sigma_{1}^{p}+\sigma_{1}^{h}$ can be recast in a more concise configuration
\begin{eqnarray}
\rho(k)&=&\frac{1}{2 \pi}+\cos k  \int_{-A}^{A} \mathrm{d} \Lambda a_{1}(\Lambda-\sin k)\sigma_{1}(\Lambda),\label{rho} \\
\sigma_{1}(\Lambda)&=&- \int_{-A}^{A} \mathrm{d} \Lambda^{\prime} a_{2}(\Lambda-\Lambda^{\prime})\sigma_{1}(\Lambda^{\prime}) +\int_{-Q}^{Q} \mathrm{d} k a_{1}(\sin k-\Lambda) \rho(k). \label{sigma}
\end{eqnarray}

In general, The momentum of paticles for $|k|\le Q, |\Lambda|\le A$ or holons if $|k|> Q, |\Lambda|> A$ with relative parameter $k_j,\Lambda_{\alpha}^{n},\Lambda_{\alpha}^{\prime n}$ can be calculated through
\begin{equation}
\begin{aligned}
&p(k)=\frac{2\pi  I_k}{L}=2\pi\int_0^k\mathrm{d}k^{\prime}\rho\left(k^{\prime}\right),\\&p_1(\Lambda)=\frac{2\pi J_1}{L}=2\pi\int_0^{\Lambda}\mathrm{d}\Lambda^{\prime}\sigma_1\left(\Lambda^{\prime}\right),\\
&p_n(\Lambda)=\frac{2\pi J_n}{L}=2\pi\int_0^{\Lambda}\mathrm{d}\Lambda^{\prime}\sigma_n^h\left(\Lambda^{\prime}\right),\ n\geq2,\\&p_n^{\prime}(\Lambda)=\frac{2\pi J_n^{\prime}}{L}=-2\pi\int_0^{\Lambda}\mathrm{d}\Lambda^{\prime}\sigma_n^{\prime h}\left(\Lambda^{\prime}\right)+\pi(n+1),\ n\geq1,\label{q}
\end{aligned}
\end{equation}
which will be used to study elementary excitations.


The aim of this Supplementary Materials is to provide derivations of the key results which are for the first time to be reported in our main text, i.e. analytical results of 
spin incoherent Luttinger liquid, asymptotic of correlation functions,  interaction-driven criticality, as well as the lattice Contact susceptibilities and quantum cooling.

\section{Phase diagram}

By the fact that thermodynamic properties feature dramatic fluctuations around quantum critical point (QCP), the ground state phase diagram can be characterized by dimensionless parameters grouped with thermodynamic quantities at low temperature. 
One dimensionless quantity is Wilson ratio
\begin{equation}R^{\chi_s}_w=\frac{4}{3}\left(\frac{\pi k_B}{\mu_Bg}\right)^2\frac{\chi_s}{C_v/T},\end{equation} 
which characterizes the nature of the TLLs, remarkably map out the full phase diagram of the 1D repulsive Hubbard model, consisting of five phases, see the Fig.1 in main text. A significant aspect of these ratio is that it is approximate constant away from the critical point. In Ref.~\cite{LPG:preparation}, we analytically calculate relevant thermodynamic quantities in all TLL and quantum criticality regions. The numerical simulation in Fig.1 of main text and analytical results in Ref.~\cite{LPG:preparation} both show that the Wilson ratio in these five phases have the values of
\begin{eqnarray}
\text{II:}\quad R^{\chi_s}_w&\approx& 2,\\\text{IV:}\quad R^{\chi_s}_w&>& 2,\\\text{V:}\quad R^{\chi_s}_w&>& 4,\\\text{I and III:}\quad R^{\chi_s}_w&=& 0.
\end{eqnarray}

Moreover, in terms of bosonization results of the magnetic susceptibility, specific heat,  Luttinger parameter and velocity,  the Wilson ratios are given by 
\begin{eqnarray}
\text{II:}\quad R^{\chi_s}_w&\approx& 2,\\\text{IV:}\quad R^{\chi_s}_w&\approx&2v_cK_s/(v_s+v_c),\label{R4-B}\\\text{V:}\quad R^{\chi_s}_w&\approx& 4K_s,\\\text{I and III:}\quad R^{\chi_s}_w&=& 0,
\end{eqnarray}
where $K_s$ is the Luttinger parameter for spin, and  $v_{c,s}$ are sound velocities for charge and spin, respectively.  
However, the $K_s$ cannot be completely determined at finite magnetic field using bosonination. 
From bosonination theory, we only know that spin Luttinger parameter $K_s>1$ for repulsive Hubbard model at finite magnetic field. Therefore, the numerical results are in agreement with the bosonination theory. 

On the other hand, Wilson ratio gives rise to a dimensionless scaling function in the quantum critical region, see our later calculation with the eqs.(\ref{1-2})-(\ref{4-5}): 
\begin{equation}
R^{\chi_s}_w\approx \frac{2\pi^2\alpha^2_B}{3}\frac{\mathrm{Li}_{-\frac{1}{2}}\left(-\mathrm{e}^x\right)}{\frac{3}{4} \mathrm{Li}_{\frac{3}{2}}\left(-\mathrm{e}^x\right)- x\mathrm{Li}_{\frac{1}{2}}\left(-\mathrm{e}^x\right) + x^2\mathrm{Li}_{-\frac{1}{2}}\left(-\mathrm{e}^x\right)}, \label{R_QCP}
\end{equation}
where the meaning of $\alpha_B,x$ can be seen from eq.(\ref{cv}). 
The spin-incoherent Luttinger liquid  will  be shown to occur  near a  transition of from phase IV to II in dilute limit,  i.e. a crossover  region between TLL phase boundary and the critical temperature in the $T-u$ or $T-\mu$ plane.
The value of Wilson ratio is  in a range  between the ones determined by the eq.(\ref{R4-B}) and eq.(\ref{R_QCP}).

\subsection{Two-holon excitations and antiholon-spinon excitations}
Flipping one spin-down particle upon the ground state will result in two-spinon fractional excitation. Together with particle-hole excitation of charge, these excitation modes exhibit spin-charge separation with quantum numbers $\left(\Delta\eta^z,\Delta S^z\right)=(0,1)$. Therefore, each of the two spinons carry no charge but spin-$\frac{1}{2}$.
 Meanwhile, in the charge sector, one particle can be removed from inside to outside the Fermi surface, corresponding to particle-hole excitation with quantum numbers $\left(\Delta\eta^z,\Delta S^z\right)=(0,0)$.
In the long wavelength limit, i.e. $\Delta K\to 0$, both charge and spin excitations exhibit linear dispersion: $ \Delta E_{c,s} = v_{c,s} \hbar |\Delta K|$. 
Hence, the slopes of such spin and charge low-energy excitations determine the velocities of the spin and charge density waves, respectively.
The energies and momenta for these two two-parametric excitations are 
\begin{eqnarray}
E_{ph}&=&\kappa(k_p)-\kappa(k_h),\quad P_{ph}=p(k_p)-p(k_h), \\
E_{ss}&=&-\varepsilon_1(\Lambda_{h1})-\varepsilon_1(\Lambda_{h2}),\quad P_{ss}=-p_1(\Lambda_{h1})-p_1(\Lambda_{h2})+\pi n_c,
\end{eqnarray}
where $k_p,\, k_h$ are the positions of particle and hole in charge degrees of freedom, $\Lambda_{h1},\,\Lambda_{h2}$ denote two hole positions in the spin sector, $\pi n_c$ comes from the parity change. 
To obtain these two-parametric spectra, firstly fix one of the components at a certain value,  and plot the single particle excitation corresponding to the other component. Then let the fixed value change, and repeat the previous step to draw a single particle excitation for another quantity.
 Repeating this procedure again and agin until finally the values of every  components are combined to obtain a complete excitation spectra in the first Brillouin zone, see Fig.2 in the main text.
 In such elementary excitaions, spin and charge modes are fully separated and characterised by  different $v_c,v_s$.
  Especially, in the low-density limit, spin sector is highly suppressed.
   Thus at  certain  finite temperatures,  one situation can occur: the temperature is larger than the characteristic energy scale of spin $E_s\sim k_Fv_s$, but still lower than typical energy scale of charge $E_c\sim k_Fv_s$,
   such that the collective excitation still remains in the charge sector, but no such collective motion in the spin sector, as reflected in Fig.2 in the main text.
Here the temperature scale is beyond the typical energy scale in spin, rendering the spin as highly excited state.
 This case is termed as the spin incoherent Luttinger liquid (SILL).
 A comprehensive description of the SILL has been theoretically studied in terms of bosonization and field theory, see \cite{Fiete:2007,Cheianov:2004}.


Let us consider the antiholon-spinon  excitation by adding  a particle with spin up or down over  the ground state.
 These two cases are classified by the charge and spin magnetization values $(\Delta \eta^z, \Delta S^z)=(\frac{1}{2},-\frac{1}{2})$
and $(\frac{1}{2},\frac{1}{2})$, indicating the emergence of fractional excitations of charge and spin originated from the discrete symmetries characterised by the spin flip and Shiba transformation \cite{Ess05}. 
Note that existence of charge fractional excitation is unique in lattice models and absent in continuous systems.  
 With the help of eqs.(\ref{q}), the excitation spectra are given by 
\begin{eqnarray}
E_{\overline{h}\overline{s}}&=&\kappa(k_p)+\varepsilon_1(\Lambda_{p}),\quad P_{\overline{h}\overline{s}}=p(k_p)+p_1(\Lambda_p)+\pi n_c,\\
E_{\overline{h}s}&=&\kappa(k_p)-\varepsilon_1(\Lambda_{h}),\quad P_{\overline{h}s}=p(k_p)-p_1(\Lambda_h),
\end{eqnarray}
where $k_p$ is the quasimomentum  of the added particle in charge degree, $\Lambda_{p},\Lambda_{h}$ denote particle and hole quasimomenta  in spin sector, $\pi n_c$ comes from the parity change. 
We draw the excitation spectra following the same as the method  used  for obtaining  the two-spinon excitation.
We observe that these excitations correspond to spin-charge scattering state, which strongly coupled together and outside of the TLL regime. It is of importance to experimentally measure such excitations via photoemission experiment like Angle Resolved Photoemission Spectroscopy.

\section{Spin-charge separation in low temperature}
Under the condition of $B/T\gg 1$ and $|\mu|/T\gg 1$, the TBA equations can be simplified to the following  coupled equations for real $k$  and length-$1$ $\Lambda$ string 
\begin{eqnarray}
\kappa(k)&=&-2 \cos{k}-\mu-2 u-B-\int_{-\infty}^{\infty}\mathrm{d} \Lambda  a_1(\sin{k}-\Lambda) T\ln(1+\mathrm{e}^{-\frac{\varepsilon_1(\Lambda)}{T}})\label{kl0},\\
\varepsilon_1(\Lambda)&=&2 B -\int_{-\pi}^{\pi} \mathrm{d} k \cos k a_1(\sin{k}-\Lambda) T\ln(1+\mathrm{e}^{-\frac{\kappa(k)}{T}})\nonumber \\
&&+\int_{-\infty}^{\infty} \mathrm{d} \Lambda^{\prime} a_{2}\left(\Lambda-\Lambda^{\prime}\right) T \ln(1+\mathrm{e}^{-\frac{\varepsilon_1(\Lambda^{\prime})}{T}}),\label{el0}
\end{eqnarray}
and the free energy is given by
\begin{equation}
f=-\int_{-\pi}^{\pi} \frac{\mathrm{d} k}{2\pi} T\ln(1+\mathrm{e}^{-\frac{\kappa(k)}{T}})+u\label{f}.
\end{equation}
By observing the structure of the expressions, the main challenge is the simplification of the integral terms, and one way to deal with these integrals is using Sommerfeld expansion. 
Denote the fermi point of $\kappa(k),\varepsilon(\Lambda)$ as $Q,A$ which satisfy $\kappa(Q)=0,\varepsilon(A)=0$. Define $\frac{\partial{\bar{a}_1(\sin k,\Lambda)}}{\partial{\Lambda}}=a_1(\sin k-\Lambda)$, $\frac{\partial{\bar{a}(\sin k,\Lambda)}}{\partial{\Lambda}}=a_1(\sin k-\Lambda)+a_1(\sin k+\Lambda)$ and the second term in eq.(\ref{kl0}) is given by 
\begin{eqnarray}
&&-\int_{-\infty}^{\infty}\mathrm{d} \Lambda  a_1(\sin{k}-\Lambda) T\ln(1+\mathrm{e}^{-\frac{\varepsilon_1(\Lambda)}{T}})\nonumber \\
&=&-\bar{a}_1(\sin k,\Lambda) T\ln(1+\mathrm{e}^{-\frac{\varepsilon_1(\Lambda)}{T}})|_{-\infty}^{\infty}-\int_{-\infty}^{\infty}\mathrm{d} \Lambda  \bar{a}_1(\sin{k},\Lambda)\frac{1}{1+\mathrm{e}^{\frac{\varepsilon_1(\Lambda)}{T}}}\frac{\varepsilon_1(\Lambda)}{\partial{\Lambda}}\nonumber\\
&=&-\int_{0}^{\infty}\mathrm{d} \Lambda  \bar{a}(\sin{k},\Lambda)\frac{1}{1+\mathrm{e}^{\frac{\varepsilon_1(\Lambda)}{T}}}\frac{\varepsilon_1(\Lambda)}{\partial{\Lambda}}\nonumber\\
&=&-\int_{\varepsilon_1(0)}^{\varepsilon_1(\infty)}\mathrm{d} \varepsilon_1 \bar{a}(\sin{k},\Lambda(\varepsilon_1))\frac{1}{1+\mathrm{e}^{\frac{\varepsilon_1}{T}}}\nonumber \\
&=&-\int_{\frac{\varepsilon_1(0)}{T}}^{\frac{\varepsilon_1(A)}{T}}\mathrm{d}x T \bar{a}(\sin{k},\Lambda(Tx))\frac{1}{1+\mathrm{e}^{x}}-\int_{\frac{\varepsilon_1(A)}{T}}^{\frac{\varepsilon_1(\infty)}{T}}\mathrm{d}x T \bar{a}(\sin{k},\Lambda(Tx))\frac{1}{1+\mathrm{e}^{x}}\nonumber \\
&=&-\int_{\frac{\varepsilon_1(0)}{T}}^{\frac{\varepsilon_1(A)}{T}}\mathrm{d}x T \bar{a}(\sin{k},\Lambda(Tx))\left(1-\frac{1}{1+\mathrm{e}^{-x}}\right)-\int_{\frac{\varepsilon_1(A)}{T}}^{\frac{\varepsilon_1(\infty)}{T}}\mathrm{d}x T \bar{a}(\sin{k},\Lambda(Tx))\frac{1}{1+\mathrm{e}^{x}}\nonumber \\
&=& -\int_{\frac{\varepsilon_1(0)}{T}}^{0}\mathrm{d}x T \bar{a}(\sin{k},\Lambda(Tx))+\int_{0}^{-\frac{\varepsilon_1(0)}{T}}\mathrm{d}x T \bar{a}(\sin{k},\Lambda(-Tx))\frac{1}{1+\mathrm{e}^{x}}-\int_{0}^{\frac{\varepsilon_1(\infty)}{T}}\mathrm{d}x T \bar{a}(\sin{k},\Lambda(Tx))\frac{1}{1+\mathrm{e}^{x}}\nonumber 
\\
&=&-\int_{\varepsilon_1(0)}^{0}\mathrm{d} \varepsilon_1 \bar{a}(\sin{k},\Lambda(\varepsilon_1))+\int_{0}^{\infty}\mathrm{d}x T \frac{\bar{a}(\sin{k},\Lambda(-Tx))-\bar{a}(\sin{k},\Lambda(Tx))}{1+\mathrm{e}^{x}}\nonumber \\
&=&\int_{\varepsilon_1(0)}^{0}\mathrm{d} \varepsilon_1 \frac{\partial{\Lambda}}{\partial{\varepsilon_1}} \varepsilon_1 (a_1(\sin{k}-\Lambda)+a_1(\sin{k}+\Lambda))-\int_{0}^{\infty}\mathrm{d}x T \frac{(a_1(\sin{k}-\Lambda(0))+a_1(\sin{k}+\Lambda(0)))\Lambda^{'}(0)2Tx}{1+\mathrm{e}^{x}}\nonumber \\
&=&\int_{0}^{A}\mathrm{d} \Lambda \varepsilon_1(\Lambda) (a_1(\sin{k}-\Lambda)+a_1(\sin{k}+\Lambda))-2T^2(a_1(\sin{k}-A)+a_1(\sin{k}+A))\Lambda^{'}(0)\frac{\pi^2}{12}\nonumber \\
&=&\int_{0}^{A}\mathrm{d} \Lambda \varepsilon_1(\Lambda) (a_1(\sin{k}-\Lambda)+a_1(\sin{k}+\Lambda))-\frac{\pi^2T^2}{6}\frac{1}{\varepsilon_1^{'}(A)}(a_1(\sin{k}-A)+a_1(\sin{k}+A)).
\end{eqnarray}
In the above calculations, the following equations are used:
\begin{eqnarray}
&& \frac{\bar{a}(\sin{k},\Lambda(Tx))-\bar{a}(\sin{k},\Lambda(-Tx))}{2Tx}=\frac{\partial{\bar{a}}}{\partial{\Lambda}}|_{\Lambda=\Lambda(0)}\Lambda^{'}(0),\nonumber \\
&& \Lambda^{'}(0)=\frac{1}{\frac{d\varepsilon_1}{d\Lambda}|_{\varepsilon_1=0}}=\frac{1}{\varepsilon^{'}_1(A)},\qquad 
\int_{0}^{\infty}\mathrm{d} x \frac{x}{1+\mathrm{e}^{x}}=\frac{\pi^2}{12}.
\end{eqnarray}

Conducting  similar manipulations in the integrals of $\varepsilon_1(\Lambda)$, the dressed energies can be converted to
\begin{eqnarray}
\kappa(k)&=&-2 \cos{k}-\mu-2 u-B\nonumber\\&&+\int_{-A}^{A}\mathrm{d} \Lambda \varepsilon_1(\Lambda) a_1(\sin{k}-\Lambda)-\frac{\pi^2T^2}{6}\frac{1}{\varepsilon_1^{'}(A)}(a_1(\sin{k}-A)+a_1(\sin{k}+A)), \label{kl0_m}\\
\varepsilon_1(\Lambda)&=&2 B- \frac{\pi^2T^2}{6\kappa^{'}(Q)}\cos Q (a_1(\sin{Q}-\Lambda)+a_1(\sin{Q}+\Lambda))+\frac{\pi^2T^2}{6\varepsilon_1^{'}(A)} (a_2(\Lambda-A)+a_2(\Lambda+A))\nonumber \\
&&+\int_{-Q}^{Q}\mathrm{d} k \kappa(k) \cos k a_1(\sin{k}-\Lambda)-\int_{0}^{A}\mathrm{d} \Lambda^{'} \varepsilon_1(\Lambda^{'}) (a_2(\Lambda-\Lambda^{'})+a_2(\Lambda+\Lambda^{'}))\label{el0_m}.
\end{eqnarray}

Comparing  above equations with root densities (\ref{rho}), (\ref{sigma}), we gain insight that there is an implicit connection between the two sets of equations which are highly symmetric. In order to derive a compact and closed form, combine eq.(\ref{rho}), (\ref{sigma}) with (\ref{kl0_m}), (\ref{el0_m}), the free energy can be associated with velocities:
\begin{eqnarray}
f&=&-\int_{-\pi}^{\pi} \frac{\mathrm{d} k}{2\pi} T\ln(1+\mathrm{e}^{-\frac{\kappa(k)}{T}})+u\nonumber-\frac{\pi^2T^2}{6\kappa^{'}(Q)}+\int_{0}^{Q} \frac{\mathrm{d} k}{\pi}\kappa(k)+u\nonumber\\
&=& -\frac{\pi^2T^2}{6\kappa^{'}(Q)}+\int_{-Q}^{Q} \mathrm{d}k\left[-2 \cos{k}-\mu-2 u-B-\frac{\pi^2T^2}{6}\frac{1}{\varepsilon_1^{'}(A)}(a_1(\sin{k}-A)+a_1(\sin{k}+A))\right]\rho(k)+u\nonumber\\
&&+\int_{-A}^{A} \mathrm{d}\Lambda \left[2 B- \frac{\pi^2T^2}{6\kappa^{'}(Q)}\cos Q (a_1(\sin{Q}-\Lambda)+a_1(\sin{Q}+\Lambda))+\frac{\pi^2T^2}{6\varepsilon_1^{'}(A)} (a_2(\Lambda-A)+a_2(\Lambda+A))\right]\sigma_{1}(\Lambda)\nonumber\\
&=&-\frac{\pi^2T^2}{6\kappa^{'}(Q)}+f_0-\frac{\pi^2T^2}{6\varepsilon_1^{'}(A)}\left[\sigma_{1}(A)+\sigma_{1}(-A)+\int_{-A}^{A} \mathrm{d}\Lambda(a_2(\Lambda-A)+a_2(\Lambda+A))\sigma_{1}(\Lambda)\right]\nonumber\\
&&-\frac{\pi^2T^2}{6\kappa^{'}(Q)}\left[\rho(Q)+\rho(-Q)-\frac{1}{\pi}\right]+\frac{\pi^2T^2}{6\varepsilon_1^{'}(A)}\int_{-A}^{A} \mathrm{d}\Lambda(a_2(\Lambda-A)+a_2(\Lambda+A))\sigma_{1}(\Lambda)\nonumber\\
&=&f_0-\frac{\pi^2T^2\rho(Q)}{3\kappa^{'}(Q)}-\frac{\pi^2T^2\sigma_{1}(A)}{3\varepsilon_1^{'}(A)}\nonumber\\&=&f_0-\frac{\pi T^2}{6}\left(\frac{1}{v_c}+\frac{1}{v_s}\right),\label{eq:f}
\end{eqnarray}
where $f_0=\int_{-Q}^{Q} \mathrm{d}k\left(-2 \cos{k}-\mu-2 u-B\right)\rho(k)+2B\int_{-A}^{A} \mathrm{d}\lambda \sigma_1(\lambda)+u$ is the background contribution from ground state. $v_c,v_s$ denote the velocities of charge and spin by  the definition $v_c=\frac{\kappa^{'}(Q)}{2\pi\rho(Q)},v_s=\frac{\varepsilon_1^{'}(A)}{2\pi\sigma_1(A)}$. 
Specific heat is directly derived $C_v=\frac{\pi T }{3}\left(\frac{1}{v_c}+\frac{1}{v_s}\right)$.
 In spin polarized band II, spin degree vanishes. 
 The free energy at low temperature is simplified as $f=f_0-\frac{\pi T^2}{6}\frac{1}{v_c}$ with specific heat $C_v=\frac{\pi T }{3}\frac{1}{v_c}$ accordingly. 
 While in phase V with charge half filled, the free energy is denoted as $f=f_0-\frac{\pi T^2}{6}\frac{1}{v_s}$, specific heat $C_v=\frac{\pi T}{3}\frac{1}{v_s}$.

\section{Universal scaling functions near quantum criticality point}
At zero temperature, the phase transition occurs when certain degrees of freedom appears, disappears or reaches saturation. 
At finite temperature, these quantum criticality span into the V-shaped critical region. 
In this section we show that the free energies at quantum criticality share the following universal forms:
\begin{eqnarray}
\text{I-II: }f&=&u+T^{\frac{3}{2}}\pi^{\frac{1}{2}}\rho(0)\left(\frac{\kappa^{''}(0)}{2}\right)^{-\frac{1}{2}}\mathrm{Li}_{\frac{3}{2}}\left(-\mathrm{e}^{-\frac{\kappa(0)}{T}}\right), \label{1-2}\\
\text{II-III: }f&=&f_0+T^{\frac{3}{2}}\pi^{\frac{1}{2}}\rho(\pi)\left(\frac{-\kappa^{''}(\pi)}{2}\right)^{-\frac{1}{2}}\mathrm{Li}_{\frac{3}{2}}\left(-\mathrm{e}^{\frac{\kappa(\pi)}{T}}\right), \label{2-3}\\
\text{V-III: }f&=&f_0+T^{\frac{3}{2}}\pi^{\frac{1}{2}}\sigma_1(0)\left(\frac{\varepsilon^{''}_1(0)}{2}\right)^{-\frac{1}{2}}\mathrm{Li}_{\frac{3}{2}}\left(-\mathrm{e}^{-\frac{\varepsilon_1(0)}{T}}\right), \label{3-5}\\
\text{II-IV: }f&=&f_0-\frac{\pi T^2}{6v_c}+T^{\frac{3}{2}}\pi^{\frac{1}{2}}\sigma_1(0)\left(\frac{\varepsilon^{''}_1(0)}{2}\right)^{-\frac{1}{2}}\mathrm{Li}_{\frac{3}{2}}\left(-\mathrm{e}^{-\frac{\varepsilon_1(0)}{T}}\right), \label{2-4}\\
\text{V-IV: }f&=&f_0-\frac{\pi T^2}{6v_s}+T^{\frac{3}{2}}\pi^{\frac{1}{2}}\rho(\pi)\left(\frac{-\kappa^{''}(\pi)}{2}\right)^{-\frac{1}{2}}\mathrm{Li}_{\frac{3}{2}}\left(-\mathrm{e}^{\frac{\kappa(\pi)}{T}}\right), \label{4-5}
\end{eqnarray}
where $f_0$ comes from the ground state, the terms with $T^2$ reflect the properties of TLL for noncritical degrees of freedom, $\sigma_1(0)$ denotes the spin density at $\Lambda=0$, $\varepsilon^{''}_1(0) \equiv \left. \frac{d^2\varepsilon_1}{d\Lambda^2} \right|_{\Lambda=0}$,  $\rho(0),\rho(\pi)$ denotes the charge density at $k=0,\pi$, and $\kappa^{''}(0) \equiv \left. \frac{d^2\kappa}{dk^2} \right|_{k=0},\kappa^{''}(\pi) \equiv \left. \frac{d^2\kappa}{dk^2} \right|_{k=\pi}$. 
Special function $\mathrm{Li}_n$ denotes the polylog function. In this note, we mainly concentrate on eqs.(\ref{2-4}), (\ref{4-5}). 
The functions  $-\varepsilon_1(0),-\kappa(0),\kappa(\pi)$ are independent of temperature in the polylog function $\mathrm{Li}_{\frac{3}{2}}$.
They depend on the energy gaps away from the QCPs, i.e.  
\begin{eqnarray}
-\varepsilon_1(0), -\kappa(0), \kappa(\pi)&\approx&\alpha_B\Delta B+\alpha_\mu\Delta \mu+\alpha_u\Delta u,
\end{eqnarray}
where $\Delta B=B-B_c,\Delta \mu=\mu-\mu_c,\Delta u=u-u_c$ denote the distances away from QCPs $(B_c,\mu_c,u_c)$; 
The factors $\alpha_{(B,\mu,u)}$ in front of $\Delta (B,\mu,u)$ represent the paths across the QCPs.
These expressions eqs.(\ref{1-2})-(\ref{4-5}) are universal scaling forms for the second order phase transitions associated with the critical dynamical experiment $z=2$ and correlation length exponent $\nu=1/2$. 
The coefficients $\alpha_{B,\mu,u}$ can be derived from the free energy and the low temperature TBA equations. 
 Here we just give the proof for eqs.(\ref{2-4}), (\ref{4-5}). 

\subsection{The transition of IV-II:}
In the vicinity of the phase transition from phase IV to phase II, the spin degrees of freedom vanish gradually, and play a leading role at criticality. 
Thus we can safely expand the kernels $a_n(\sin k-\Lambda),\, a_n(\Lambda^{'}-\Lambda)$ around the point of $\Lambda=0$ in TBA eqs.(\ref{kappa}), (\ref{epsilon}) to get the terms in the powers of $\Lambda^n$.
 This procedure decouples the convolution terms.  Firstly, $a_1(\sin k-\Lambda)$ expands as 
\begin{equation}
a_1(\sin k-\Lambda)\approx -\frac{A_1(k)}{2}-\frac{A_2(k)}{2}\Lambda^2+O(\Lambda^4), 
\end{equation}
where $A_1(k)=-2a_1(\sin k),\, A_2(k)=2\pi a_1^2(\sin k)/u-8\pi^2\sin^2 ka_1^3(\sin k)/u^2$. 
We keep the orders up to $O(\Lambda^2)$ in whole calculations. Applying this expansion form  into eqs.(\ref{kl0}), (\ref{el0}), we get the leading terms
\begin{eqnarray}
\kappa(k)&=&-2\cos k-\mu-2u-B+A_1(k)I_1+A_2(k) I_2, \label{k24}\\
\varepsilon_1(\Lambda)&=&D_1\Lambda^2+D_2,
\end{eqnarray}
where $I_1,I_2$ are integrals term with respect to $\varepsilon_1(\Lambda)$, which denoted by
\begin{eqnarray}
I_1&=&\int_{0}^{\infty} \mathrm{d}\Lambda  T \ln(1+\mathrm{e}^{-\frac{\varepsilon_1(\Lambda)}{T}})=-\frac{T^{\frac{3}{2}}\pi^{\frac{1}{2}}}{2D^{\frac{1}{2}}_1}\mathrm{Li}_{\frac{3}{2}}\left(-\mathrm{e}^{\frac{-D_2}{T}}\right), \\
I_2&=&\int_{0}^{\infty} \mathrm{d} \Lambda \Lambda^2 T \ln(1+\mathrm{e}^{-\frac{\varepsilon_1(\Lambda)}{T}})=-\frac{T^{\frac{5}{2}}\pi^{\frac{1}{2}}}{4D^{\frac{3}{2}}_1}\mathrm{Li}_{\frac{5}{2}}\left(-\mathrm{e}^{\frac{-D_2}{T}}\right),
\end{eqnarray}
and $D_1,D_2$ are functions of $\kappa(k),I_1,I_2$ (explicit expressions in Ref.~\cite{LPG:preparation}). In the limit of $\Lambda \rightarrow0$, the spin density at $\Lambda=0$ has simple form $\sigma(0)=\int_{-Q}^{Q} \mathrm{d} k/(2\pi)a_1(\sin k)$.  Conducting Sommerfeld expansion in free energy eq.(\ref{f}) firstly, and then substituting eq.(\ref{k24}) into free energy eq.(\ref{f}), we have
\begin{eqnarray}
f&=&-\frac{\pi T^2}{6\kappa^{'}(Q)}+\int_{0}^{Q} \frac{\mathrm{d} k}{\pi}\kappa(k)+u\\&&=-\frac{\pi T^2}{6\kappa^{'}(Q)}+f_0+I_1\int_{0}^{Q} \frac{\mathrm{d} k}{\pi}A_1(k)\\&&=-\frac{\pi T^2}{6\kappa^{'}(Q)}+f_0-2\sigma_1(0)I_1\\&&=f_0-\frac{\pi T^2}{6v_c}+T^{\frac{3}{2}}\pi^{\frac{1}{2}}\sigma_1(0)\left(\frac{\varepsilon^{''}_1(0)}{2}\right)^{-\frac{1}{2}}\mathrm{Li}_{\frac{3}{2}}\left(-\mathrm{e}^{-\frac{\varepsilon_1(0)}{T}}\right), 
\end{eqnarray}
where $f_0=\int_{0}^{Q} \mathrm{d} k/\pi\left(-2 \cos{k}-\mu-2 u-B\right)+u$ and $D_1=\varepsilon^{''}_1(0)/2,D_2=\varepsilon_1(0)$  were used. 
By the analysis of critical behavior, the  argument function $\varepsilon_1(0)$  in the polylog functions can be deduced from the TBA equations,
\begin{equation}
\varepsilon_1(0)=2B-\int_{-\pi}^{\pi}\mathrm{d}k \cos ka_1(\sin k-\Lambda)T\ln(1+\mathrm{e}^{-\frac{\kappa(k)}{T}})|_{\Lambda=0}.
\end{equation}

The phase transition occurs at $\varepsilon_1(0)=0$ with charge expressed as $\kappa(k)=-2\cos k-\mu-2u-B$. Let $Q$ denote  the Fermi point of $\kappa(k)$, and the  boundary condition satisfies $2\cos Q=-\mu-2u-B$. Utilizing  this boundary condition at  zero temperature, $\varepsilon_1(0)$ has the following form
\begin{equation}
\varepsilon_1(0)=2B-\frac{4u}{\pi}\int_{0}^{Q}\mathrm{d}k\frac{\cos k}{u^2+\sin^2k}(\cos k-\cos Q)+O(T^2). 
\end{equation}
By further  performing  expansion around the critical  points $(B_c,\mu_c,u_c)$,  one can give  
\begin{equation}
\cos Q\approx\cos Q_c-\sin Q_c\left[(\partial Q/\partial B)|_{Q=Q_c}\Delta B+(\partial Q/\partial \mu)|_{Q=Q_c}\Delta \mu+(\partial Q/\partial u)|_{Q=Q_c}\Delta u\right],
\end{equation}
 with $Q_c=\arccos\left(-\frac{1}{2}(\mu_c+2u_c+B_c)\right)$. Through some algebraic calculations, $\alpha_B,\alpha_{\mu},\alpha_u$ can thus be obtained for the transition from II to IV
\begin{eqnarray}
\alpha_B&=&-2\left[1-\frac{1}{\pi}\arctan\left(\frac{\sin Q_c}{u}\right)\right], \\
\alpha_{\mu}&=&\frac{2}{\pi}\arctan\left(\frac{\sin Q_c}{u}\right), \\
\alpha_u&=&\frac{4}{\pi}\left[\int_{0}^{Q_c}\mathrm{d}k\frac{\cos^2 k}{u_c^2+\sin^2 k}-2u_c^2\int_{0}^{Q_c}\mathrm{d}k\frac{\cos^2 k}{(u_c^2+\sin^2 k)^2}\right. \nonumber\\
&&+\left.\arctan\left(\frac{\sin Q_c}{u_c}\right)+\frac{\sin(2Q_c)}{2(u_c^2+\sin Q_c^2)}\right]. \label{u}
\end{eqnarray}
We observe from above equations that $\alpha_B,\alpha_{\mu}$ can be directly calculated, although a closed form of $\alpha_u$ cannot be found. 
We note that the interaction strength $u$ appears in the kernel of the integral, leading to a significant challenge in the calculation of quantum scaling functions.
Nevertheless at   quantum criticality, the heat capacity can be obtained through the second partial of the free energy with respect to temperature
\begin{equation}
C_v/T = c_0+c_1T^{-1/2} \left[ \frac{3}{4} \mathrm{Li}_{\frac{3}{2}}\left(-\mathrm{e}^x\right)- x\mathrm{Li}_{\frac{1}{2}}\left(-\mathrm{e}^x\right) + x^2\mathrm{Li}_{-\frac{1}{2}}\left(-\mathrm{e}^x\right)\right], \label{cv}
\end{equation}
where in the transition of phase IV to II, $c_0=\pi/(3v_c)$ presents the contribution from the  background, $c_1=-\pi^{\frac{1}{2}}\sigma_1(0)(\varepsilon^{''}_1(0)/2)^{-1/2}$, and $x=-\varepsilon_1(0)/T=(\alpha_B \Delta B+\alpha_\mu\Delta \mu+\alpha_u\Delta u)/T$.
 By numerical simulations, we observe that  specific heat displays a bimodal structure, whose local maxima mark the critical crossover temperatures fanning out from the critical point. 
 The local maxima can be determined by the condition $\partial C_v/\partial B(\mu,u)=0$, leading the condition 
\begin{equation}
\frac{1}{4} \mathrm{Li}_{\frac{1}{2}}\left(-\mathrm{e}^x\right)- x\mathrm{Li}_{-\frac{1}{2}}\left(-\mathrm{e}^x\right) - x^2\mathrm{Li}_{-\frac{3}{2}}\left(-\mathrm{e}^x\right)=0,
\end{equation}
that  gives two solutions $x_1=-1.5629,\,x_2=3.6205$. 
We also observe that  the analytic result (\ref{cv})   agrees well with the numerical results, showing that  our estimations are rather efficient.
Fig.3 in the main text shows the contour plot of specific heat in the temperature-interaction strength plane. In this critical region, $T\gg \Delta u= u-u_c$, thermodynamic properties of the system  can be cast into  universal scaling forms, for  example, see eq.(\ref{cv}).
 
The yellow line with square symbols in Fig.3 in the main text represents the boundary of the TLL region, which resides below the line, characterized by a specific heat $C_v$ linearly dependent on $T$. To the left of $u_c=1$, the TLL contains only one component. To the right of $u_c$, the TLL contains both the spin and the charge degrees of freedom. 
In this regime, spin-charge separation \cite{Haldane:1981,Recati:PhysRevLett.90.020401,Guan:2012,Mestyan:2019spin,Patu:PhysRevB.101.035149} can be observed. 
Above the TLL$_C$ phase the spin sector is gapped, utilizing the asymptotic behaviour of polylog function\cite{mohankumar2007two} and expanding the results from QC part eq.(\ref{cv}), the specific heat is given by
\begin{equation}
C_v\approx \frac{\pi T}{3v_c}+\frac{3\pi^{\frac{1}{2}}}{4} \sigma_1(0)\left(\frac{\varepsilon^{''}_1(0)}{2}\right)^{-\frac{1}{2}}T^{\frac{1}{2}}\mathrm{e}^{\alpha_u\Delta u/T}+O\left(\Delta u\mathrm{e}^{\alpha_u\Delta u/T}\right). 
\end{equation}
 
In contrast, for the temperature $E_s\sim k_Fv_s\ll  k_BT\ll E_c\sim k_Fv_c$  above the TLL$_{SC}$ phase boundary,  the specific heat is given by 
\begin{equation}
C_v\approx \frac{\pi T}{3v_c}+\frac{\pi^2\sigma_1(0)\left(\varepsilon^{''}_1(0)/2\right)^{-\frac{1}{2}}(-\varepsilon_1(0))^{-\frac{1}{2}}T}{3}\left[1+\frac{21\pi^2}{40}(-\varepsilon_1(0))^{-2}T^2\right]+O(T^4),
\end{equation}
which showing a gas-liquid co-existence.

\subsection{Correlation function in SILL}

In the TLL regime $T\ll E_c,E_s$, the finite size corrections to the energy of low-lying excited states in terms of the changes of particle numbers, backward scattering, particle and hole  are expressed by  \cite{Ess05,Giamarchi:book,Guan:2013}
\begin{equation}
\Delta E=\frac{2\pi}{L}\sum_{\alpha={\rm 
		c,s}}v_\alpha (N_\alpha^-+N_\alpha^+)+\frac{\pi}{2L}
\sum_{\alpha={\rm c,s}} \left\{v_\alpha(\hat Z^{-1}\Delta
\vec N)_\alpha^2+v_\alpha[\hat Z^{\rm t} (2\vec D)]_\alpha^2\right\}, 
\end{equation}
where $v_\alpha$ are Fermi velocities, $N_\alpha^\pm,\Delta
\vec N,\vec D$ are related to the three types of excitations: adding particle at left and right Fermi points, change total particle number, backward scattering, and $\hat Z$ denotes dressed charge matrix \cite{Ess05}. 
In contrast, the bosonization Hamiltonian is given by \cite{Giamarchi:book,Guan:2013}
\begin{eqnarray}
H&=& \sum_{\alpha={\rm c,s}}\sum_{q\neq0}v_\alpha |q| \hat 
b^\dag_{\alpha,q} \hat b_{\alpha,q}~+ \frac{\pi}{2L}\sum_{\alpha={\rm 
		c,s}} \left(~v^\alpha_{\rm N} \Delta N_\alpha^2 ~+~v^\alpha_{\rm J}J_\alpha^2 ~\right)\\
\nonumber
&& +\frac{2c}{L}\sum_{k_1,k_2,p}\sum_{r=\pm1}\hat 
\psi^{\dag}_{\uparrow,r,k_1} \hat \psi^\dag_{\downarrow,-r,k_2} \hat 
\psi_{\downarrow,r,k_2+2rk_{\rm F}+p}
\hat\psi_{\uparrow,-r,k_1-2rk_{\rm F}-p},
\end{eqnarray}
where $\hat b^\dag \,(\hat b)$ denotes creation (annihilation) operator of an boson with quasimomentuem $q$, $v^\alpha_{\rm N},v^\alpha_{\rm J}$ denotes density stiffness and phase stiffness, $c$ is relevant to coupling strength, $J_\alpha$ is the current operator and $\hat \psi$ the Fermi field operator. Based on finite-size correction, the two-point correlation functions of the primary field at finite temperature are expressed in terms of conformal dimensions 
\begin{eqnarray}
\left\langle \phi(x,t)\phi(0,0)\right\rangle _T&=&\sum A(D_c,D_s,N^{\pm}_c,N^{\pm}_s)\text{exp}(-2\mathrm{i}D_ck_{F,\uparrow}x)\text{exp}(-2\mathrm{i}(D_c+D_s)k_{F,\downarrow}x)\nonumber\\&&\times \left(\frac{\pi T}{v_c\text{sinh}(\pi T(x-\mathrm{i}v_ct)/{v_c})}\right)^{2\Delta^{+}_c}\left(\frac{\pi T}{v_c\text{sinh}(\pi T(x+\mathrm{i}v_ct)/{v_c})}\right)^{2\Delta^{-}_c}\nonumber\\&&\times \left(\frac{\pi T}{v_s\text{sinh}(\pi T(x-\mathrm{i}v_st)/{v_s})}\right)^{2\Delta^{+}_s}\left(\frac{\pi T}{v_s\text{sinh}(\pi T(x+\mathrm{i}v_st)/{v_s})}\right)^{2\Delta^{-}_s},\label{cft}
\end{eqnarray}
where $\Delta_c^{\pm},\Delta_s^{\pm}$ are the conformal dimensions associated with elementary excitations and dressed charge matrix $\hat Z$ given in Ref.~\cite{Ess05}.

However, in the SILL regime, the spin excitation is suppressed severely due to the temperature scale  $T>E_s$, and hence the spin sector is non-dynamic, while charge maintains relevant low-energy behaviour. 
Although the TLL theory holds under the condition $T\ll E_c,E_s$, 
we can separately treat the energy scales of charge and spin degrees of freedom in the SILL regime, also see Fig.~3 in the main text. 
For the temperature $E_s\ll T\ll E_c$ \cite{Cheianov:2004}, the  finite-temperature correlation functions of the conformal field theory eq.(\ref{cft}) still remain valid for the charge and spin  degrees of freedom under different conditions
\begin{eqnarray}
 |x\pm \mathrm{i}v_ct|\ll v_c/T,\qquad  |x\pm \mathrm{i}v_st|\gg v_s/T,\label{condition}
 \end{eqnarray}
 which is essential to capture the asymptotic behaviour of the SILL. 
 Here we remark that the typical energy scale of spin can be given by $E_s\sim J\sim \left(k_{F\uparrow}+k_{F\downarrow}\right)/2\cdot v_s\equiv k_Fv_s$, where $J$ stands for the effective exchange coupling of spin chain induced by interaction \cite{Guan:2013}. 
 Thus under the condition (\ref{condition}),   the single-particle Green's function and pair correlation function are formally given by 
 \begin{eqnarray}
 G^{\uparrow }&\! \approx \! & \mathrm{e}^{-\mathrm{i} k_{F,\uparrow} x}{\cal C}_{\uparrow}^-\left( x-\mathrm{i} v_c t\right) \langle S_R^+\left(x,t\right)S_R(0,0)\rangle
 +h.c., \label{Correlation-Green}\\
 G^{p}&\! \approx \! & \mathrm{e}^{-\mathrm{i} \left( k_{F,\uparrow} + k_{F,\downarrow} \right) x}{\cal C}_{p}^-\left( x-\mathrm{i} v_c t\right) 
 {\cal C}_{p}^+\left( x+\mathrm{i} v_c t\right) \langle S_R^+\left(x,t\right)S_R(0,0)\rangle +h.c.,  \label{Correlation-pair}
 \end{eqnarray}
where the charge correlations ${\cal C}_{\uparrow}^-\left(Z\right)\sim  1/Z^{2\Delta_c^+}$, ${\cal C}_{p}^{\pm} \left(Z\right)\sim  1/Z^{2\Delta_c^{\mp}}$ decay as a power-law of distance, whereas the spin mode correlation $\langle S_R^+\left(x,t\right)S_R(0,0)\rangle\sim \left(2\pi \alpha k_F\right)^{2\Delta^{+}_s+2\Delta^{-}_s}\mathrm{e}^{-\pi\alpha\left(2\Delta^{+}_s+2\Delta^{-}_s\right) k_Fx}$ decays exponentially. Here $\alpha$ is a constant to be determined. 
In the limit of high density and low spin-down density, the conformal dimensions can be given in  the order no more than $O(\hat{n}_c,n_{\downarrow})$ \cite{LPG:preparation}
\begin{eqnarray}
G^{\uparrow}&:&\,2\Delta^+_c\approx0,\,\,2\Delta^-_c\approx1-\zeta_1n_{\downarrow},\,\,2\Delta^+_s\approx\frac{1}{4}-\frac{1}{u}\hat{n}_c-\frac{3}{4}\zeta_1n_{\downarrow},2\Delta^-_s\approx\frac{1}{4}-\frac{1}{u}\hat{n}_c+\frac{1}{4}\zeta_1n_{\downarrow};\\
G^{p}&:&\, 2\Delta^{+}_c\approx\frac{9}{4}-\frac{3}{u}\hat{n}_c,\,\,2\Delta^{-}_c\approx\frac{1}{4}-\frac{1}{u}\hat{n}_c,\nonumber \\
&& 2\Delta^{+}_s\approx\frac{1}{4}+\frac{1}{u}\hat{n}_c-\frac{3}{4}\zeta_1n_{\downarrow},\,\,2\Delta^{-}_s\approx\frac{1}{4}-\frac{1}{u}\hat{n}_c-\frac{3}{4}\zeta_1n_{\downarrow},
\end{eqnarray}
where $\hat n_c=1-n,\zeta_1=\sqrt{1+u^2}/u$ with $n,n_{\downarrow}$ particle number and down-spin number.
For the special case $B=0$, we can compare the Luttinger parameter determined from the Wilson ratio and  the conformal dimensions are given in Ref.~\cite{Ess05},
\begin{equation}
G^{\uparrow}:\,2\Delta^+_c\approx\frac{1}{16},\,\,2\Delta^-_c\approx\frac{9}{16},\,\,2\Delta^+_s\approx0,2\Delta^-_s\approx\frac{1}{2}\label{B=0},\end{equation}
and observe that these results eq.(\ref{B=0}) agree with those given in Ref.~\cite{Cheianov:2004}, corresponding to the case of spinless fermions with Luttinger parameter $K_c=1/2$.
Although the bosonization expressions of the conformal dimensions were   already given in \cite{Cheianov:2004,Fiete:2007},
here  our results  give new insights into such novel SILL in the 1D Hubbard model from the Bethe ansatz perspective.

\subsection{The transition of IV-V:}
In the phase transition of IV-V, the charge is gradually saturated as $\kappa(\pi)$ approaching zero.
 Thus we expand the kernels $a_n(\sin k-\Lambda)$ around the point of $\sin k=0$ in TBA eqs.(\ref{kappa}) and (\ref{epsilon}) in  terms of the power of $\sin^nk$. Then charge dispersion can be written as  the functions of $\cos k$ and $\sin k$. We first expand the kernal function $a_1(\sin k-\Lambda)$  in the form 
\begin{equation}
a_1(\sin k-\Lambda)\approx a_1(\Lambda)-(\pi a^2_1(\Lambda)/u-4\pi^2\Lambda^2 a^3_1(\Lambda)/u^2)\sin^2k+O(\sin^4k).
\end{equation}
Applying this expansion form into eqs.(\ref{kl0}), (\ref{el0}), we have the result 
\begin{eqnarray}
\kappa(k)&=&-2\cos k+2C_1\sin^2k+C_2, \label{k45}\\
\varepsilon_1(\Lambda)&=&2B-\int_{-\pi}^{\pi} \mathrm{d}k 2\cos^2ka_1(\sin k-\Lambda)+\int_{-\infty}^{\infty} \mathrm{d} \Lambda^{\prime} a_{2}\left(\Lambda-\Lambda^{\prime}\right) T \ln(1+\mathrm{e}^{-\frac{\varepsilon_1(\Lambda^{\prime})}{T}})\nonumber\\&&-2J_1a_1(\Lambda)+2J_2\left[\frac{\pi}{u}a^2_1(\Lambda)-\frac{4\pi^2}{u^2}\Lambda^2a^3_1(\Lambda)\right], \label{e45}
\end{eqnarray}
where $C_1,C_2$ are integral  terms with respect to $\varepsilon_1(\Lambda)$
\begin{eqnarray} C_1&=&\int_{0}^{\infty}\mathrm{d} \Lambda \left[\frac{\pi}{u}a^2_1(\Lambda)-\frac{4\pi^2}{u^2}\Lambda^2a^3_1(\Lambda)\right] T\ln(1+\mathrm{e}^{-\frac{\varepsilon_1(\Lambda)}{T}}),\\C_2&=&-\mu-2 u-B-\int_{0}^{\infty}\mathrm{d} \Lambda  2a_1(\Lambda) T\ln(1+\mathrm{e}^{-\frac{\varepsilon_1(\Lambda)}{T}}).
\end{eqnarray}
And $J_1,J_2$ are integral terms  with respect to $\kappa(k)$ and can be calculated by integration by parts:
\begin{eqnarray}
J_1&=&T\int_{0}^{\pi} \mathrm{d} k \cos k \ln(1+\mathrm{e}^{\frac{\kappa(k)}{T}})=\frac{T^{\frac{3}{2}}}{\sqrt{1-2C_1}}\Gamma(\frac{3}{2})\mathrm{Li}_{\frac{3}{2}}\left(-\mathrm{e}^{\frac{2+C_2}{T}}\right)-\frac{T^{\frac{5}{2}}}{8(1-2C_1)^{\frac{5}{2}}}\Gamma(\frac{5}{2})\mathrm{Li}_{\frac{5}{2}}\left(-\mathrm{e}^{\frac{2+C_2}{T}}\right),\\J_2&=&T\int_{0}^{\pi} \mathrm{d} k \cos k \sin^2 k \ln(1+\mathrm{e}^{\frac{\kappa(k)}{T}})=\frac{T^{\frac{5}{2}}}{3(1-2C_1)^{\frac{3}{2}}}\Gamma(\frac{5}{2})\mathrm{Li}_{\frac{5}{2}}\left(-\mathrm{e}^{\frac{2+C_2}{T}}\right).
\end{eqnarray}

The second integral in eq.(\ref{e45}) can be obtained by the  Sommerfeld expansion, i.e. 
\begin{equation}
\int_{-\infty}^{\infty} \mathrm{d} \Lambda^{\prime} a_{2}\left(\Lambda-\Lambda^{\prime}\right) T \ln(1+\mathrm{e}^{-\frac{\varepsilon_1(\Lambda^{\prime})}{T}})=-\int_{-A}^{A} \mathrm{d} \Lambda^{\prime} a_{2}\left(\Lambda-\Lambda^{\prime}\right)\varepsilon_1(\Lambda^{\prime})+\frac{\pi^2T^2}{6\varepsilon_1^{'}(A)} (a_2(\Lambda-A)+a_2(\Lambda+A)).
\end{equation}

On the other hand, the density functions in the limit of $k\rightarrow\pi$ have the form
\begin{eqnarray}
\rho(\pi)&=&\frac{1}{2 \pi}- \int_{-A}^{A} \mathrm{d} \Lambda a_{1}(\Lambda)\sigma_{1}(\Lambda),\label{rho45} \\
\sigma_{1}(\Lambda)&=&- \int_{-A}^{A} \mathrm{d} \Lambda^{\prime} a_{2}(\Lambda-\Lambda^{\prime})\sigma_{1}(\Lambda^{\prime}) +\frac{1}{2 \pi}\int_{-\pi}^{\pi} \mathrm{d} k a_{1}(\sin k-\Lambda).\label{sigma45}
\end{eqnarray}
Substituting   the expressions of densities eq.(\ref{rho45}), (\ref{sigma45}) into free energy, then we calculate  the contributions from ground state  near Mott phase 
\begin{eqnarray}
f_0&=&\int_{-\pi}^{\pi} \mathrm{d}k\left(-2 \cos{k}-\mu-2 u-B\right)\rho(k)+2B\int_{-A}^{A} \mathrm{d}\lambda \sigma_1(\lambda)+u\nonumber\\&=&-\mu-u-B+\int_{-\pi}^{\pi} \mathrm{d}k\int_{-A}^{A} \mathrm{d}\lambda \sigma_1(\lambda)\cos k(-2\cos k-\mu-2u-B)a_1(\sin k-\Lambda)+2B\int_{-A}^{A} \mathrm{d}\lambda \sigma_1(\lambda)\nonumber\\&=&-\mu-u-B-\int_{-A}^{A} \mathrm{d}\lambda \sigma_1(\lambda)\int_{-\pi}^{\pi} \mathrm{d}k 2\cos^2 ka_1(\sin k-\Lambda)+2B\int_{-A}^{A} \mathrm{d}\lambda \sigma_1(\lambda)\nonumber\\&=&-\mu-u-B+\int_{-A}^{A} \mathrm{d}\lambda \sigma_1(\lambda)\varepsilon^0(\Lambda), 
\end{eqnarray}
where $\varepsilon^0(\Lambda)=2B-\int_{-\pi}^{\pi} \mathrm{d}k 2\cos^2ka_1(\sin k-\Lambda)$. Therefore the free energy at low temperature via integration by part is given by
\begin{eqnarray}
f&=&-T \int_{-\pi}^{\pi} \frac{\mathrm{d} k}{2 \pi} \ln \left(1+\mathrm{e}^{\frac{-\kappa(k)}{T}}\right)+u\nonumber\\&=&\int_{-\pi}^{\pi} \frac{\mathrm{d} k}{2\pi}\kappa(k)+u-T \int_{-\pi}^{\pi} \frac{\mathrm{d} k}{2 \pi} \ln \left(1+\mathrm{e}^{\frac{\kappa(k)}{T}}\right)\nonumber\\&=&\int_{-\pi}^{\pi} \frac{\mathrm{d} k}{2\pi}\kappa(k)+u+\frac{1}{\pi}\int_{0}^{\pi}\mathrm{d}k\frac{(k-\pi)}{1+\mathrm{e}^{-\frac{\kappa(k)}{T}}}\frac{\partial\kappa(k)}{\partial k}-\frac{T}{\pi}(k-\pi) \ln \left(1+\mathrm{e}^{\frac{\kappa(k)}{T}}\right)|_{0}^{\pi}\nonumber\\&=&\int_{-\pi}^{\pi} \frac{\mathrm{d} k}{2\pi}\kappa(k)+u-T\ln(1+\mathrm{e}^{\frac{\kappa(0)}{T}})+\frac{J_1}{\pi}+\frac{J_2}{2\pi}.\label{f45}
\end{eqnarray}

Using eq.(\ref{kl0}), the first integral in eq.(\ref{f45}) can be written as
\begin{eqnarray}
&&\int_{-\pi}^{\pi} \frac{\mathrm{d} k}{2\pi}\kappa(k)\nonumber= -\mu-2u-B-\int_{-\pi}^{\pi} \frac{\mathrm{d} k}{2\pi} \int_{-\infty}^{\infty}\mathrm{d}\Lambda a_1(\sin{k}-\Lambda) T\ln(1+\mathrm{e}^{-\frac{\varepsilon_1(\Lambda)}{T}})\nonumber\\&=&-\mu-2u-B+\int_{-\pi}^{\pi} \frac{\mathrm{d} k}{2\pi}\left\{\int_{-A}^{A}\mathrm{d}\Lambda a_1(\sin{k}-\Lambda)\varepsilon_1(\Lambda)-\frac{\pi^2T^2}{6\varepsilon_1^{'}(A)} (a_1(\sin k-A)+a_2(\sin k+A))\right\},
\end{eqnarray}
where the second term can be dealt with through the density eq.(\ref{sigma45})
\begin{eqnarray}
&&\int_{-\pi}^{\pi} \frac{\mathrm{d} k}{2\pi}\int_{-A}^{A}\mathrm{d}\Lambda a_1(\sin{k}-\Lambda)\varepsilon_1(\Lambda)\nonumber\\&=&\int_{-A}^{A}\mathrm{d}\Lambda\varepsilon_1(\Lambda)\left[\sigma_1(\Lambda)+ \int_{-A}^{A} \mathrm{d} \Lambda^{\prime} a_{2}(\Lambda-\Lambda^{\prime})\sigma_{1}(\Lambda^{\prime})\right]\nonumber\\&=&\int_{-A}^{A}\mathrm{d}\Lambda\sigma_1(\Lambda)\left[\varepsilon_1(\Lambda)+ \int_{-A}^{A} \mathrm{d} \Lambda^{\prime} a_{2}(\Lambda-\Lambda^{\prime})\varepsilon_{1}(\Lambda^{\prime})\right]\nonumber\\&=&\int_{-A}^{A}\mathrm{d}\Lambda\sigma_1(\Lambda)\left[\varepsilon^0(\Lambda)-2J_1a_1(\Lambda)+\frac{\pi^2T^2}{6\varepsilon_1^{'}(A)} (a_2(\Lambda-A)+a_2(\Lambda+A))\right]\nonumber\\&=&f_0+(\mu+u+B)+2J_1\left(\rho(\pi)-\frac{1}{2\pi}\right)+\frac{\pi^2T^2}{6\varepsilon_1^{'}(A)}\int_{-A}^{A}\mathrm{d}\Lambda\sigma_1(\Lambda)(a_2(\Lambda-A)+a_2(\Lambda+A)), 
\end{eqnarray}
where we ignore the terms related to $J_2$ in  the  subleading order to $J_1$. 
Thus the free energy finally can be given by  
\begin{eqnarray}
f&=&f_0+2J_1\left(\rho(\pi)-\frac{1}{2\pi}\right)+\frac{J_1}{\pi}\nonumber\\&&+\frac{\pi^2T^2}{6\varepsilon_1^{'}(A)}\left\{\int_{-A}^{A}\mathrm{d}\Lambda\sigma_1(\Lambda)(a_2(\Lambda-A)+a_2(\Lambda+A))-\int_{-\pi}^{\pi}\frac{\mathrm{d}k}{2\pi}(a_1(\sin k-A)+a_1(\sin k+A))\right\}\nonumber\\&=&f_0+2\rho(\pi)J_1-\frac{\pi^2T^2\sigma(A)}{3\varepsilon_1^{'}(A)}\nonumber\\&=&f_0-\frac{\pi T^2}{6v_s}+T^{\frac{3}{2}}\pi^{\frac{1}{2}}\rho(\pi)\left(\frac{-\kappa^{''}(\pi)}{2}\right)^{-\frac{1}{2}}\mathrm{Li}_{\frac{3}{2}}\left(-\mathrm{e}^{\frac{\kappa(\pi)}{T}}\right), 
\end{eqnarray}
where the term $\kappa(\pi)$ is given by 
\begin{equation}
\kappa(\pi)=2-\mu-2u-B+\int_{-A}^{A}\mathrm{d} \Lambda  a_1(\Lambda) \varepsilon_1(\Lambda).\label{kpi45}
\end{equation}
Here, $\varepsilon_1(\Lambda)$ is given by 
\begin{equation}
\varepsilon_1(\Lambda)=2B-\int_{-\pi}^{\pi} \mathrm{d}k 2\cos^2ka_1(\sin k-\Lambda)-\int_{-A}^{A} \mathrm{d} \Lambda^{\prime} a_{2}(\Lambda-\Lambda^{\prime})\varepsilon_1(\Lambda^{\prime}).\label{e045}
\end{equation}

It is indicated from above two equations (\ref{kpi45}), (\ref{e045}) that 
\begin{equation}
\alpha_{\mu}=-1\label{45mu},
\end{equation}
which  only appears in the leading term of charge.
 However, it's hard to derive explicit expressions of $\alpha_B$ and $\alpha_u$ due to the coupling of the two degrees of freedom. We will give  another method to solve this problem in next section.

\section{Contact susceptibility}
In quantum many-body systems, interaction plays the central role. 
In analogy to the  Contact for quantum atomic gases, we define the lattice version of the Contact $C=\partial f/ \partial u$, i.e. the derivative of the free energy with respect to the interaction.
 We observe that the Contact remarkably marks the phase boundaries. This can be seen in Fig.~\ref{c}, where we use the contour plot of the Contact to re-draw the phase diagram as in Fig.1 in the  main text. 
The two dark orange dashed lines represent the Contact $C=0$ and $C=-0.65$, showing a sudden change at phase boundaries (black dashed lines). Here, we define  the Contact susceptibilities with respect to the external potentials. 
Using the Maxwell relations, i.e., the derivative orders are commutative $\frac{\partial}{\partial T}\left(\frac{\partial f}{\partial u}\right)=\frac{\partial}{\partial u}\left(\frac{\partial f}{\partial T}\right)$,$\frac{\partial}{\partial B}\left(\frac{\partial f}{\partial u}\right)=\frac{\partial}{\partial u}\left(\frac{\partial f}{\partial B}\right)$,$\frac{\partial}{\partial \mu}\left(\frac{\partial f}{\partial u}\right)=\frac{\partial}{\partial u}\left(\frac{\partial f}{\partial \mu}\right)$, we may build up general relations between Contact susceptibilities and interaction-driven variations of density, magnetization and entropy:
\begin{eqnarray}
	\frac{\partial s}{\partial u}&=&-\frac{\partial C}{\partial T} \label{uT}, \\
	\frac{\partial n}{\partial u}&=&-\frac{\partial C}{\partial \mu}, \label{umu} \\
	\frac{\partial m}{\partial u}&=&-\frac{\partial C}{\partial (2B)}. \label{uB}
	\end{eqnarray}
These three relations relate interaction-induced phase transitions with magnetic-/chemical-induced phase transitions, and can be used to determine $\alpha_u$ appeared in scaling functions.

\begin{figure}[t] 
	\begin{center} 
		\includegraphics[width=0.8\linewidth]{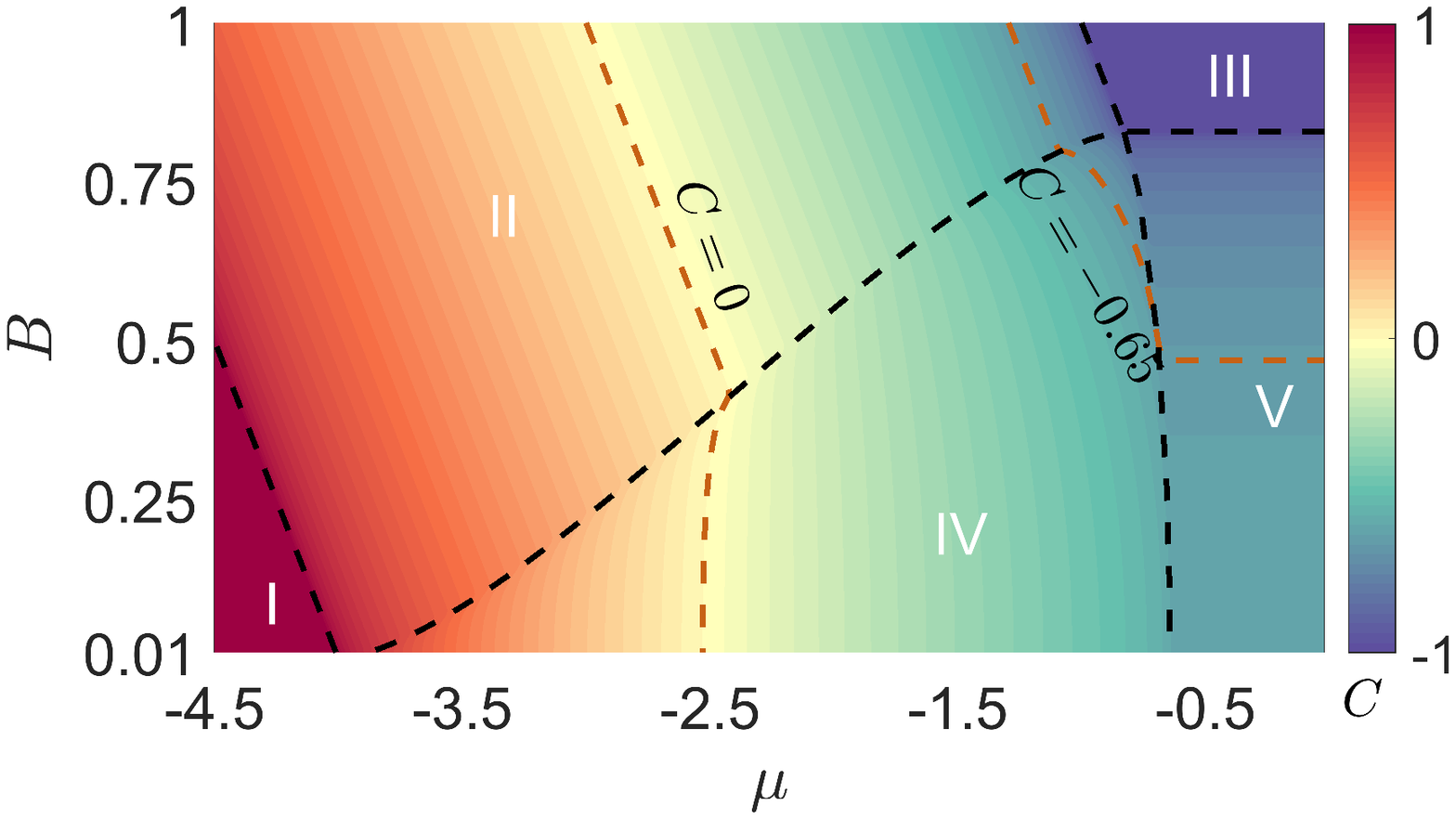} 
	\end{center} 	
	\caption{Contour plot of the Contact in $B-\mu$ plane for $T=0.005$ and $u=1$. The Contact displays a sudden changes at phase transitions, consistent with the analytical phase boundaries (black dashed lines at  zero temperature, see Fig.1 in the main text.). The dark orange dashed lines mark the contours of $C=0$ and $C=-0.65$.}       
	\label{c}     
\end{figure}

\subsection{Quantum cooling}
It is precisely the first equality eq.(\ref{uT}) of Maxwell relations that allows us to drive interaction to refrigeration isentropically. Consider the entropy as functions of $s=s(u,T)$ for fixed magnetic field,  and thus the total derivative is 
\begin{equation}\mathrm{d}s=\frac{\partial s}{\partial u}\mathrm{d}u+\frac{\partial s}{\partial T}\mathrm{d}T=0.
\end{equation}
Substitute eq.(\ref{uT}) into above expression and using $\frac{\partial s}{\partial T}=\frac{ C_v}{T}$ with $C_v$ is specific heat, the  points on the isentrope line in the $(u,T)$ coordinates admit
 \begin{equation}\frac{C_v}{T} \frac{\partial T}{\partial u}=\frac{\partial C}{\partial T}, 
 \end{equation} 
 where $C_v$ is given by eq.(\ref{cv}), $\partial C/\partial T$ can be derived with the help of eq.(\ref{2-4}), (\ref{4-5}), for example for II-V:
\begin{equation}
\frac{\partial C}{\partial T}=T^{-\frac{1}{2}}\alpha_u\pi^{\frac{1}{2}}\sigma_1(0)\left(\frac{\varepsilon^{''}_1(0)}{2}\right)^{-\frac{1}{2}}\left[\frac{1}{2}\mathrm{Li}_{\frac{1}{2}}\left(-\mathrm{e}^{x}\right)-x\mathrm{Li}_{-\frac{1}{2}}\left(-\mathrm{e}^{x}\right)\right], 
\end{equation}
where  $x=\alpha_u \Delta u/T$.
Near a critical point, by the fact that local maximum of the entropy leads to a local temperature minimum in an isentropic process, and using the condition $\frac{\partial C}{\partial T}=0$, we have 
\begin{equation} \frac{1}{2}\mathrm{Li}_{\frac{1}{2}}\left(-\mathrm{e}^{x}\right)-\frac{\alpha_u\Delta u}{T}\mathrm{Li}_{-\frac{1}{2}}\left(-\mathrm{e}^{x}\right)=0.
\end{equation} 
The approximate solution of the quantity $x\approx1.3117$. The entropy for criticality regions can be obtained from the derivatives of eqs.(\ref{2-4}), (\ref{4-5}). And the minimum temperature can be obtained, see the main text.

\subsection{Calculation of $\alpha_u$ in the transition IV-V}

In the discussion of the phase transition of IV-V,  the coefficient $\alpha_u$ associated with $\Delta u$ in the argument of the scaling function is important to determine the scaling functions of physical properties with respect to the interaction strength.
Meanwhile, $\alpha_u$ also determines the lowest temperature which can be reached in an interaction-driven quantum cooling process, i.e. 
$x_0=\alpha_u \Delta u/T\approx1.3117 $. 
Here we present a proposal of numerical estimation with the help of the eq.(\ref{umu}). 
Note that the demarcation line between phases IV and V is described by constant density $n=1$. 
Considering the chemical potential $\mu$ and interaction $u$ driven phase transitions, we conduct a total derivative of $n$
\begin{equation}
\mathrm{d}n=\frac{\partial n}{\partial u}\mathrm{d}u+\frac{\partial n}{\partial \mu}\mathrm{d}\mu=0.
\end{equation}
Substitute eq.(\ref{umu}) into above expression and using $\frac{\partial n}{\partial \mu}=\chi_c$ with $\chi_c$  compressibility, the phase points on the constant density line $n=1$ in the $(u,\mu)$ coordinates have
\begin{equation}\chi_c \frac{\partial \mu}{\partial u}=\frac{\partial C}{\partial \mu}, \label{dn}
\end{equation}
where $\chi_c,\, \partial C/\partial \mu$ are near the critical points and therefore can be obtained through the derivatives of free energy eqs.(\ref{4-5})
\begin{eqnarray}
\chi_c&=&-T^{-\frac{1}{2}}\alpha^2_{\mu}\pi^{\frac{1}{2}}\rho(\pi)\left(\frac{-\kappa^{''}(\pi)}{2}\right)^{-\frac{1}{2}}\mathrm{Li}_{-\frac{1}{2}}\left(-\mathrm{e}^{\frac{\kappa(\pi)}{T}}\right),\\
\frac{\partial C}{\partial \mu}&=&T^{-\frac{1}{2}}\alpha_{\mu}\alpha_u\pi^{\frac{1}{2}}\rho(\pi)\left(\frac{-\kappa^{''}(\pi)}{2}\right)^{-\frac{1}{2}}\mathrm{Li}_{-\frac{1}{2}}\left(-\mathrm{e}^{\frac{\kappa(\pi)}{T}}\right). 
\end{eqnarray}
Substituting these expression of $\chi_c$ and $\partial C/\partial \mu$ into eq.(\ref{dn}), we get 
\begin{equation}
\alpha_u=-\alpha_{\mu}\frac{\partial \mu}{\partial u}=\frac{\partial \mu}{\partial u},\label{a}
\end{equation}
in which we have used the identity $\alpha_{\mu}=-1$, see eq.(\ref{45mu}). 
It is expected from eq.(\ref{a}) that the slope of the crossover line in the $(u,\mu)$ coordinates for the Mott transition can evaluate the coefficient $\alpha_u(u_c,\mu_c)$. 
In view of the relation (\ref{a}),  we plot the contour diagram of  density at fixed magnetic field $B_c=0.82714$ around $u_c=1,\mu_c=-0.8272$, seen Fig.4 (b1) in the main text. 
Thus a series of transition points $(u_c,\mu_c)$ in the $(u,\mu)$ coordinate are identified at which the density firstly equals unity $n=1$. By artificially choosing two adjacent points $(u_1,\mu_1),(u_2,\mu_2)$ near $u_c=1$, then  we approximately get differential $(\mu_1-\mu_2)/(u_1-u_2)\approx-1.9627$. 
Therefore we have $\alpha_u(u_c=1)\approx-1.9627$. Using this value, we plot the  compressibility near the phase transition  driven by interaction, see Fig.4(b2) in the main text. A good agreement between the scaling function of the compressibility with this analytical result of the  $\alpha_u$ and numerical  result  from  the TBA result is seen in Fig.4(b2) in the main text. 
The relation (\ref{a}) reveals deep insights into  the transition from IV to V, i.e.  the emergence of Mott insulator with constant density and compressibility.

\subsection{Calculation of $\alpha_u$ in the transition IV-II}
In analogy with the application of eq.(\ref{umu}), the relation eq.(\ref{uB}) is also possible to derive the coefficient $\alpha_u$  associated with $\Delta u$  in the transition from II to V. 
This relation eq.(\ref{uB}) is useful to study the phase transition with a sudden change of the density of state in  spin-down electrons. 
The total derivatives on the spin-down particle density $n_{\downarrow}$ is zero at the transition line at a constant chemical poential, i.e. 
\begin{equation}
\mathrm{d}n_{\downarrow}=\frac{\partial n_{\downarrow}}{\partial u}\mathrm{d}u+\frac{\partial n_{\downarrow}}{\partial (2B)}\mathrm{d}(2B)=0. \label{tnd}
\end{equation}
Due to the Maxwell relation eqs.(\ref{uB}) and (\ref{umu}) relate density $n$ and magnetization $m$ with interaction strength $u$, we had better convert $n_{\downarrow}$ into $n,\, m$.
 Using $n_{\downarrow}=n/2-m$, thus total derivative eq.(\ref{tnd}) is transformed into the form below
\begin{equation}
\mathrm{d}n_{\downarrow}=\left(\frac{1}{2}\frac{\partial n}{\partial u}-\frac{\partial m}{\partial u}\right)\mathrm{d}u+\left(\frac{1}{4}\frac{\partial n}{\partial B}-\frac{1}{2}\frac{\partial m}{\partial B}\right)\mathrm{d}(2B)=0. 
\end{equation}

Substituting eq.(\ref{uB}) into the above expression and using $\frac{\partial m}{\partial B}=\chi_s$ with $\chi_s$ being the spin susceptibility, the points on the constant density line of $n_{\downarrow}=0$ in the $(u,B)$ coordinates meet 
\begin{equation}
\left(\frac{1}{2}\frac{\partial n}{\partial u}+\frac{\partial C}{\partial (2B)}\right)+\left(\frac{1}{2}\frac{\partial n}{\partial B}-\chi_s\right)\frac{\partial B}{\partial u}=0. \label{tnda}
\end{equation}

Using eq.(\ref{2-4}), the analytic expressions for above derivatives are given by 
\begin{eqnarray}
\frac{\partial n}{\partial u}&=&-T^{-\frac{1}{2}}\alpha_u\alpha_{\mu}\pi^{\frac{1}{2}}\sigma_1(0)\left(\frac{\varepsilon^{''}_1(0)}{2}\right)^{-\frac{1}{2}}\mathrm{Li}_{\frac{1}{2}}\left(-\mathrm{e}^{-\frac{\varepsilon_1(0)}{T}}\right),\\
\frac{\partial C}{\partial (2B)}&=&\frac{1}{2}T^{-\frac{1}{2}}\alpha_u\alpha_{B}\pi^{\frac{1}{2}}\sigma_1(0)\left(\frac{\varepsilon^{''}_1(0)}{2}\right)^{-\frac{1}{2}}\mathrm{Li}_{\frac{1}{2}}\left(-\mathrm{e}^{-\frac{\varepsilon_1(0)}{T}}\right),\\
\frac{\partial n}{\partial B}&=&-T^{-\frac{1}{2}}\alpha_B\alpha_{\mu}\pi^{\frac{1}{2}}\sigma_1(0)\left(\frac{\varepsilon^{''}_1(0)}{2}\right)^{-\frac{1}{2}}\mathrm{Li}_{\frac{1}{2}}\left(-\mathrm{e}^{-\frac{\varepsilon_1(0)}{T}}\right),\\
\chi_s&=&-\frac{1}{2}T^{-\frac{1}{2}}\alpha^2_B\pi^{\frac{1}{2}}\sigma_1(0)\left(\frac{\varepsilon^{''}_1(0)}{2}\right)^{-\frac{1}{2}}\mathrm{Li}_{\frac{1}{2}}\left(-\mathrm{e}^{-\frac{\varepsilon_1(0)}{T}}\right).
\end{eqnarray}
Then substituting these expressions into eq.(\ref{tnda}), we derive $\alpha_u$ 
\begin{equation}
\alpha_u=-\alpha_B\frac{\partial B}{\partial u},
\end{equation}
which is dependent of $u_c,\, B_c$. 
Our analysis provides an alternative way to measure themodynamic quantities based on interaction-induced quantum transition. 
Apart from the quantum transition in terms of chemical potential and magnetic field, here we present the critical properties of the 1D Hubbard model from perspective of  the interaction.

\end{widetext} 

\end{document}